\documentclass[twocolumn]{aastex63_ionthinspace}

\usepackage{xcolor}
\usepackage{amsmath}
\usepackage{affils}
\usepackage{enumitem}

\newcommand{\HI}{\ion{H}{1}}

\newcommand{\OII}{\ion{O}{2}}
\newcommand{\OIII}{\ion{O}{3}}
\newcommand{\halpha}[0]{H$\alpha$}
\newcommand{\hbeta}[0]{H$\beta$}

\newcommand{\telec}{$T_e$}
\newcommand{\ftst}[0]{[\ion{O}{3}]$\lambda$4363\AA{}}
\newcommand{\foos}[0]{[\ion{O}{3}]$\lambda$5007\AA{}}
\newcommand{\oiidoublet}[0]{[\ion{O}{2}]$\lambda\lambda$3727,3729\AA{}}
\newcommand{\sttoB}[0]{[\ion{O}{2}]$\lambda$7320\AA{}}
\newcommand{\sttoR}[0]{[\ion{O}{2}]$\lambda$7330\AA{}}
\newcommand{\etam}[1][]{\eta_{\rm m #1}}
\newcommand{\asfh}{a_{\rm SFH}}

\newcommand{\kms}{{\rm km\ s^{-1}}}
\newcommand{\tlbmax}[0]{$\sim$2.5 Gyr}
\newcommand{\tlb}[0]{$t$}
\newcommand{\logmstar}[1][]{$\log_{10}(\rm M_\star/M_\odot)$#1}

\newcommand{\tth}{\textsuperscript{th}}

\newcommand{\iffirst}[1]{}

\newcommand{\aat}{AAT/2dF}
\newcommand{\mmt}{MMT/Hectospec}
\newcommand{\sagabg}{SAGAbg-A}

\newcommand{\umdr}{UniverseMachine DR1}
\newcommand{\tng}{TNG50}

\newcommand{\nfinal}{23258}

\newcommand{\nreference}{\rrr{1124}}

\newcommand{\paperiv}{\cite{sagaiv}}
\newcommand{\rrr}[1]{\textcolor{black}{#1}}

\defcitealias{kadofong2024}{Paper I}

\graphicspath{{./}{PaperII_figures/}}

\begin{document}
\shortauthors{Kado-Fong et al.}

\title{SAGAbg II: the Low-Mass Star-Forming Sequence Evolves Significantly Between $0.05<z<0.21$}

\author[0000-0002-0332-177X]{Erin Kado-Fong}
\affiliation{Physics Department, Yale Center for Astronomy \& Astrophysics, PO Box 208120, New Haven, CT 06520, USA}

\author[0000-0002-7007-9725]{Marla Geha}
\affiliation{Department of Astronomy, Yale University, New Haven, CT 06520, USA}

\author[0000-0002-1200-0820]{Yao-Yuan Mao}
\affiliation{Department of Physics and Astronomy, University of Utah, Salt Lake City, UT 84112, USA}

\author[0000-0002-4739-046X]{Mithi A. C. de los Reyes}
\affiliation{Department of Physics and Astronomy, Amherst College, 25 East Drive, Amherst, MA 01002}

\author[0000-0003-2229-011X]{Risa H. Wechsler}
\affiliation{Kavli Institute for Particle Astrophysics and Cosmology and Department of Physics, Stanford University, Stanford, CA 94305, USA}
\affiliation{SLAC National Accelerator Laboratory, Menlo Park, CA 94025, USA}

\author[0000-0001-6065-7483]{Benjamin Weiner}
\affiliation{Department of Astronomy and Steward Observatory, University of Arizona, Tucson, AZ 85721, USA}

\author[0000-0002-8320-2198]{Yasmeen Asali}
\affiliation{Department of Astronomy, Yale University, New Haven, CT 06520, USA}

\author[0000-0002-3204-1742]{Nitya Kallivayalil}
\affiliation{Department of Astronomy, University of Virginia, 530 McCormick Road, Charlottesville, VA 22904, USA}

\author[0000-0002-1182-3825]{Ethan O. Nadler}
\affiliation{Carnegie Observatories, 813 Santa Barbara Street, Pasadena, CA 91101, USA}
\affiliation{Department of Physics \& Astronomy, University   of Southern California, Los Angeles, CA, 90007, USA}
\affiliation{Department of Astronomy \& Astrophysics, University of California, San Diego, La Jolla, CA 92093, USA}

\author[0000-0002-9599-310X]{Erik J. Tollerud}
\affiliation{Space Telescope Science Institute, 3700 San Martin Drive, Baltimore, MD 21218, USA}

\author[0000-0001-8913-626X]{Yunchong Wang}
\affiliation{Kavli Institute for Particle Astrophysics and Cosmology and Department of Physics, Stanford University, Stanford, CA 94305, USA}

\correspondingauthor{Erin Kado-Fong} 
\email{erin.kado-fong@yale.edu}
  
  \date{\today}

\begin{abstract}
The redshift-dependent relation between galaxy 
stellar mass and star formation rate (the Star-Forming Sequence, or SFS) is a key observational yardstick for galaxy assembly.
We use the \sagabg{} sample of background galaxies from the Satellites Around Galactic Analogs (SAGA) Survey to model the low-redshift 
evolution of the low-mass SFS. 
The sample is comprised of \nfinal{} galaxies with \halpha{}-based star 
formation rates (SFRs) spanning $6<\log_{10}(\rm M_\star/[M_\odot])<10$ and
$z<0.21$ ($\tlb<2.5$ Gyr).  
Although it is common to bin or stack galaxies at $z \lesssim 0.2$ for galaxy population studies, the difference in lookback time between $z=0$ and $z=0.21$ is comparable to the time between $z=1$ to $z=2$.
We develop a model to account for both 
the physical evolution of low-mass SFS and the selection function of the SAGA survey, allowing us to disentangle redshift evolution from redshift-dependent selection effects across the \sagabg{} redshift range.  Our findings indicate 
significant evolution in the SFS over the last \tlbmax{}, with a rising normalization: $\langle {\rm SFR}({\rm M_\star=10^{8.5} M_\odot)}\rangle(z)=1.24^{+0.25}_{-0.23}\ {\rm z} -1.47^{+0.03}_{-0.03}$. 
We also identify the redshift limit at which a static SFS is ruled out at the 95\% confidence level, which is $z=0.05$ based on the precision of the SAGAbg-A sample.
Comparison with cosmological hydrodynamic simulations reveals that some contemporary simulations under-predict the recent evolution of the low-mass SFS. This demonstrates that the recent evolution of the low-mass SFS can provide new constraints on the assembly of the low-mass Universe and highlights the need for improved models in this regime.

\end{abstract}

\section{Introduction}
Scaling relations are one of the most-used and oldest tools in the extragalactic astronomer's toolkit \citep[see, e.g.][]{faber1976, tully1977, kormendy1985, kennicutt1989, kennicutt1998,pahre1998,ferrarese2001, bolton2007, canodiaz2016}. 
These mathematically simple relations are powerful summaries of the galaxy population, 
and see continued use in the field both in their own right \citep[e.g.][]{delosreyes2019, pessa2021, yao2022}, as jumping-off points for more 
complex frameworks to describe the many facets of the galaxy population \citep[see, e.g.][]{jiang2021, ostriker2022, kadofong2022c, sun2023}, 
and as benchmarks for 
simulated galaxies \citep[see, e.g.][]{brooks2007, wang2015, hopkins2018, nelson2019, munshi2021}.

The star-forming sequence (SFS), which describes the relation between star-forming 
galaxies' stellar masses and their star formation rates, has been extensively measured and analysed in the literature \citep{brinchmann2004, noeske2007, pannella2009, wuyts2011, karim2011, rodighiero2011, whitaker2012, speagle2014, whitaker2014, tomczak2016, pannella2015, renzini2015, lee2015, pearson2018, leslie2020, leja2022, dale2023}. The shape, normalization, and scatter of the SFS encode information about galaxy assembly as a function of stellar mass, how the bulk galaxy population evolves in SFR over time, and serves as a key benchmark by which cosmological simulations evaluate the realism of their simulated galaxy populations.

Observations of Nearby galaxies --- for some definition of ``Nearby'', which we will here use to refer to objects no more distant than a few tens of Mpc \citep[see, e.g.][]{dale2009, lee2009, leroy2019, dale2023} --- are contextualized by large-volume observations of the galaxy population at larger distances. \rrr{While observations of star formation in the Nearby Universe are of course important constraints on the star formation cycle due to both the precision and spatial scale that are observationally achievable at these distances \citep[see, e.g.,][]{mcquinn2015, lin2019, leroy2019, ellison2020,morselli2020, pessa2021,pessa2022, sun2023, ellison2024}, }
the large volumes of these more distant galaxy populations are important for two reasons. First, by merit of volume alone, the total size of these samples far outstrips Nearby galaxy samples. These larger samples are necessary to precisely characterize the 
distribution of galaxy properties in addition to their average properties. Second, many 
Nearby galaxy samples are environmentally biased due to their proximity to the Milky Way, M31, or the Local Group. Reaching galaxies at somewhat larger distances is necessary to 
avoid confounding environmentally driven processes with generic and/or secular processes in galaxy evolution. 

When considering the SFS and other scaling relations --- and more generally galaxy evolution --- galaxies at $z\lesssim 0.1-0.3$ are often thought of as physically equivalent to galaxies in the~$z=0$~Universe \citep[see, e.g.,][]{salim2007, peng2010, andrews2013, renzini2015, battisti2017, salim2018}. It is common practice to stack, bin, or otherwise jointly analyze galaxies from across 
this range; this approach is certainly practical in 
that it increases the size of one's ``Local analog'' sample and has been shown to be reasonable 
for areas in which systematic uncertainties or statistical uncertainties prevent a precise comparison
between Nearby and $z>0$ samples \citep[see, for example,][]{cook2014, leroy2019, dale2023}. As our view of the $z>0$ low-mass \rrr{(\logmstar[$\lesssim 9.5$])} galaxy population 
becomes more precise, however, it is increasingly important to understand how truly Nearby galaxies (those at distances less than a few tens of Mpc) compare to those at \rrr{the somewhat arbitrarily-defined ``low-redshift'' Universe.}

\rrr{Much of the present work is concerned with the physical implications of the adopted definition of ``low-redshift'', and the subtleties inherent in comparing our Nearby Universe to large-volume samples that span a larger distance (and thus lookback time) in an evolving Universe. As discussed above, it is common to take the ``$z\sim0$'' approximation; that is, the assumption that the underlying population from which the large-volume sample has been drawn is the same as the Nearby Universe in regards to redshift evolution. However, we will argue that there are three relevant redshift regimes that are often encompassed within the phrase ``low-redshift'':}

\begin{description}
  \item[\rrr{The Nearby Universe}] \rrr{Objects that are no more distant than a few tens of Mpc and including the Milky Way, for which redshift evolution is negligible (lookback times $\lesssim 100$ Myr)}
  \item[\rrr{The $z\sim0$ Universe}] \rrr{The set of objects for which redshift evolution is negligible with respect to the Nearby Universe.} 
  \item[\rrr{The $z>0$ Universe}] \rrr{The set of objects within the low-$z$ Universe wherein significant redshift evolution has rendered the underlying population of galaxies statistically distinct from the Nearby Universe.} 
\end{description}

\rrr{The numerical value of the redshift that separates the $z\sim0$ and $z>0$ samples is dependent on sample precision; this paper will establish such a demarcation for a benchmark spectroscopic survey of low-mass galaxies. In this work, we will focus on the
demarcation between $z\sim0$ and $z>0$ rather than the upper redshift limit of low-redshift, though we note that this limit is typically taken to be $z\lesssim 0.3$; we adopt an upper redshift limit of $z\sim 0.2$ for the present work..}

\rrr{Understanding the physical implications of comparisons against low-redshift galaxy samples becomes} especially important for surveys such as the Satellites Around Galactic Analogs (SAGA) Survey, a spectroscopic census of the satellite systems \rrr{of} 101 MW-like hosts \citep{geha2017, mao2021, sagaiii, sagaiv, sagav}.  A main goal of the SAGA survey 
is to put the Milky Way into cosmological context and quantify the impact of host processing on satellite galaxy properties; to do so, it is crucial to quantitatively understand how to build a fair reference sample of the ``bulk'' $z\sim 0$ galaxy population at the mass scale of MW-like satellite systems.

The sample of \textit{non-\rrr{SAGA} satellite} galaxies observed by the SAGA Survey is well-\rrr{suited} to a study of low-redshift dwarf galaxy evolution.
The SAGA Survey itself is a census of satellites down to $M_V\approx 12$ around Milky Way-like hosts in the Nearby Universe. To ensure a complete spectroscopic survey of the satellite population with photometric target selection, the SAGA Survey targeted a large number of low-mass, low-redshift galaxies. Indeed, the
vast majority of the spectra \rrr{($\gtrsim\! 99\%$; the SAGA survey obtained around 45,000 redshifts and identified 378 SAGA host satellites)} collected by the SAGA Survey are of low-mass background galaxies that are not associated with the target SAGA host.
These background 
spectra represent a sample of primarily low-mass galaxies down to a limiting magnitude of 
$m_r\sim 21$, which is more than a magnitude fainter than the Galaxy And Mass Assembly \citep[GAMA;][]{baldry2010}. 
This sample thus provides a 
fainter and deeper look into the low-redshift, low-mass galaxy population than has been 
possible with previous generation spectroscopic surveys.

In \citetalias{kadofong2024} of this series, we developed a simple model to describe the observed redshift evolution of SAGA background 
galaxies by considering the evolution of their stellar mass, star formation rate, gas-phase metallicity, and gas mass (the $M_\star$-SFR-$Z_O$-$M_g$ space)
and both the physical and observational processes that affect the 
observed properties of the galaxy sample. These processes include
their mean \HI{} gas fraction, the \rrr{strength of their galactic} winds, and the mean star formation history \rrr{averaged over the population as a function of redshift}.
In the paper at hand, we will aim to understand the impact of this mean star formation history on the evolution of the SFS and the implications of an evolving SFS on our broader attempts to understand galaxy evolution. In particular,
we quantify the \rrr{maximum} redshift at which the $z\sim 0$ \rrr{approximation holds true; i.e., the maximum redshift at which the} SFS is statistically 
indistinguishable from the $z=0$ SFS of Nearby galaxies, and
compare the redshift evolution of the observed
SFS at \logmstar[$\lesssim 9$] to observational expectations from previous work at higher mass and/or redshift, and theoretical predictions.

We adopt a flat $\Lambda$CDM cosmology with $\Omega_m=0.3$ ($\Omega_\Lambda=0.7$), and
$H_0=70\ \kms{}\ {\rm Mpc^{-1}}$. We use a \cite{kroupa2001} initial mass function (IMF) unless
otherwise specified, and convert literature results that use other IMFs to a 
\cite{kroupa2001} IMF
as stated in the text.

\section{The SAGA Background Sample}
The sample considered in this work is fully described in~\citetalias{kadofong2024}.
We will reiterate the relevant facets of the sample selection such that the reader may
understand the present work as a standalone, but direct the reader interested in 
further technical detail to~\citetalias{kadofong2024}. 

The SAGA survey is a spectroscopic search for satellites down to $M_r\sim-12$ around
MW-like hosts at $z\lesssim 0.013 $ \citep{geha2017, mao2021, sagaiii, sagaiv, sagav}.  However, the vast majority of spectra collected as part of the SAGA survey are low-redshift, low-mass background galaxies.
In this work we consider a subset of the SAGA background galaxies which, following \citetalias{kadofong2024}, we will call the \sagabg{} sample. These galaxies 
are chosen to be low-mass (\logmstar[$<10$]) out to a 
redshift of $z=0.21$; this upper redshift limit was 
determined in \citetalias{kadofong2024} as the
maximum redshift at which the observed 
wavelength of \halpha{} lies within a spectral 
region with well-estimated absolute flux calibrations.
The ``bg-A'' nomenclature refers to the fact that these galaxies are background (bg) galaxies whose spectra originate either from archival Sloan Digital Sky Survey \citep[SDSS,][]{abolfathi2018} or Galaxy And Mass Assembly Survey \citep[GAMA,][]{baldry2018} observations, or from the SAGA spectroscopic campaign on \aat{} and \mmt{}. 
The spectra from the dedicated SAGA observations account 
for 96\% of the sample.
We perform an absolute flux calibration to the SAGA and SDSS spectra 
using SAGA photometry measured from Legacy Survey imaging \citep[for details regarding SAGA photometric measurements see][]{sagaiii}. We do not apply an absolute flux calibration to the GAMA spectra because the published spectra are already flux-calibrated in a consistent manner. 

We adopt the stellar masses and \halpha{}-derived star formation rates corrected for both Galactic and internal reddening as measured in \citetalias{kadofong2024} given the spectroscopic redshift obtained by the SAGA survey. Details on the observation, spectroscopic processing, and emission line measurements of the \sagabg{} sample can be found therein; here we will summarize the most salient points.  

\subsection{Emission Line Flux Measurements}\label{s:methods:emlines}
We measure line fluxes from the \sagabg{} via a joint fit to an ensemble of 
emission lines and Balmer absorption features. The emission lines in this set are:
\halpha, \hbeta, 
\ftst, \foos{}, \oiidoublet, \sttoB, and \sttoR. Our joint modeling approach allows us to 
use physically-motivated priors for lines of the same species; we impose that 
the line ratio of any two lines originating from the same species should be within the
range of line ratios that are physically allowed for typical ISM conditions, modulo 
reddening from extinction. For example, for \halpha{} and \hbeta{} we adopt the prior:
\begin{equation}
  \rm Pr(\mathcal{A}_{H\alpha}, \mathcal{A}_{H\beta}) = (1 + e^{-k(\mathcal{A}_{H\alpha}/\mathcal{A}_{H\beta} - \tilde{\mathcal{A}}_{H\alpha,H\beta})})^{-1},
\end{equation} 
where
$k=30$ is the shape parameter of the logistic function, $\tilde{\mathcal{A}}_{H\alpha,H\beta}$ is the intrinsic
flux ratio (here assumed to be 2.86), and $\rm \mathcal{A}_{H\alpha}$ and $\rm \mathcal{A}_{H\beta}$ are the
amplitudes of the \halpha{} and \hbeta{} lines. We then estimate line fluxes for 
the full set of emission lines by adopting a Gaussian likelihood:
\begin{equation}
    \begin{split}
      \ln\mathcal{L}(\vec F_\lambda | \vec \theta_\ell) = -\frac{1}{2}\sum_{i}\left[\frac{{\left( \hat F_\lambda(\lambda_i,\vec \theta_\ell) - F_{\lambda,i} \right)}^2}{\sigma_{F_{\lambda,i}}^2} + \right. \\
      \left. \ln{\left( 2 \pi \sigma_{F_\lambda,i}^2\right)}\right],
    \end{split}
\end{equation} 
where $i$ is the index of the spectral resolution element, $\hat F_\lambda(\lambda_i)$ and $F_{\lambda,i}$ are the predicted and observed spectral flux densities in that resolution element, respectively, and $\sigma_{F_\lambda}$ is the uncertainty in those spectra flux densities. 
The observed spectrum is then $\vec F_\lambda = \{F_{\lambda,0}, F_{\lambda,1} ... F_{\lambda, M}\}$ for $M$ spectral resolution elements.
The vector $\vec \theta_\ell$ is the set of line model parameters and is of length $2N_{\rm em} + N_{\rm cont} + 3$. Here, $N_{\rm em}$ is the number of emission lines fit and $N_{\rm cont}$
is the number of continuum regions fit. Emission lines within $140\AA$ of each other are assumed to have the same continuum spectra flux density. 

Each individual line is modeled with a 
Gaussian line profile characterized by its known observed wavelength (as all of the \sagabg{} spectra have already been redshifted) and a standard deviation $\sigma_{\rm em}$. This linewidth $\sigma_{\rm em}$ is held fixed for 
all lines. The Balmer absorption features are likewise modeled with a Gaussian 
profile where the amplitude of each feature is required to be nonpositive and the 
equivalent width is held to be $\rm EW_{H\gamma} = EW_{H\delta} = EW_{H\beta} = 2EW_{H\alpha}$, following \cite{gonzalezdelgado1999}. The local continuum underneath
each line is assumed to be constant. 

The line flux measurements for the \sagabg{} sample, as well as positional and redshift measurements for the full SAGA background sample, are published in machine-readable format as part of \cite{sagaiii}. \rrr{In this work, we estimate the 3$\sigma$ limiting
\halpha{} flux of the spectrum from the standard deviation of the nearby off-line spectrum ($14 < |\lambda - \lambda_{\rm H\alpha}| < 140$ \AA{}), and remove 
galaxies wherein the median posterior \halpha{} flux is less than this limiting flux.}

\subsection{Star Formation Rates}
Star formation rates are estimated from extinction-corrected \halpha{} luminosities following \cite{calzetti2013}. \rrr{We compute \halpha{} luminosities using the luminosity distance corresponding to the catalog SAGA redshift for the source; aperture corrections are taken into account via the absolute flux calibration (see \citetalias{kadofong2024}, Section 2.2).} \rrr{In all cases, we adopt} a \cite{kroupa2001} initial mass function (IMF). That is, we assume that:
\begin{equation}
    {\rm SFR(L_{H\alpha})} = \beta_c L_{\rm H\alpha},
\end{equation}
where $\beta_c=5.5\times10^{-42} M_\odot\ \text{erg}^{-1}$.  We correct for internal extinction of \halpha{} using our estimate of $A_V$ from \citetalias{kadofong2024}; 
formally, we estimate $A_V$, \telec{}[{\OIII}], and \telec{}[\OII{}] simultaneously via a joint fit constrained by the Balmer decrement and temperature-sensitive [\OIII{}] and [\OII{}] line ratios. 
These electron temperatures were used to estimate gas-phase metallicity in \citetalias{kadofong2024}; because the wavelength range spanned by the temperature-sensitive line pairs is small compared to that spanned by the Balmer decrement, one can consider our $A_V$ measurements to be equivalent in practice to estimates of $A_V$ derived from the Balmer decrement alone.

We note that this SFR prescription differs from the \halpha{} equivalent width-based prescription used for the SAGA 
satellites themselves in \cite{sagaiv}, for which not all of the spectra
were absolutely flux calibrated, but that the two prescriptions 
agree to within measurement precision in cases where both the 
\halpha{} equivalent width and line luminosity have been measured. 

\subsection{Stellar Masses}
Stellar masses are estimated from the restframe SAGA photometry:
\begin{equation}
  \log_{10}\left( \frac{M_\star}{M_\odot} \right) = 1.254 + 1.098(g-r)_{\rm o} - 0.4 M_{r,\rm o},
\end{equation}
where the ``o'' subscript above indicates restframe measurements following \cite{sagaiii}. As in \citetalias{kadofong2024} and \cite{mao2021}, we assume an uncertainty of 0.2 dex in our stellar mass estimates.

\subsection{The Final Sample}
\begin{figure*}[t]
  \centering     
  \includegraphics[width=0.75\linewidth]{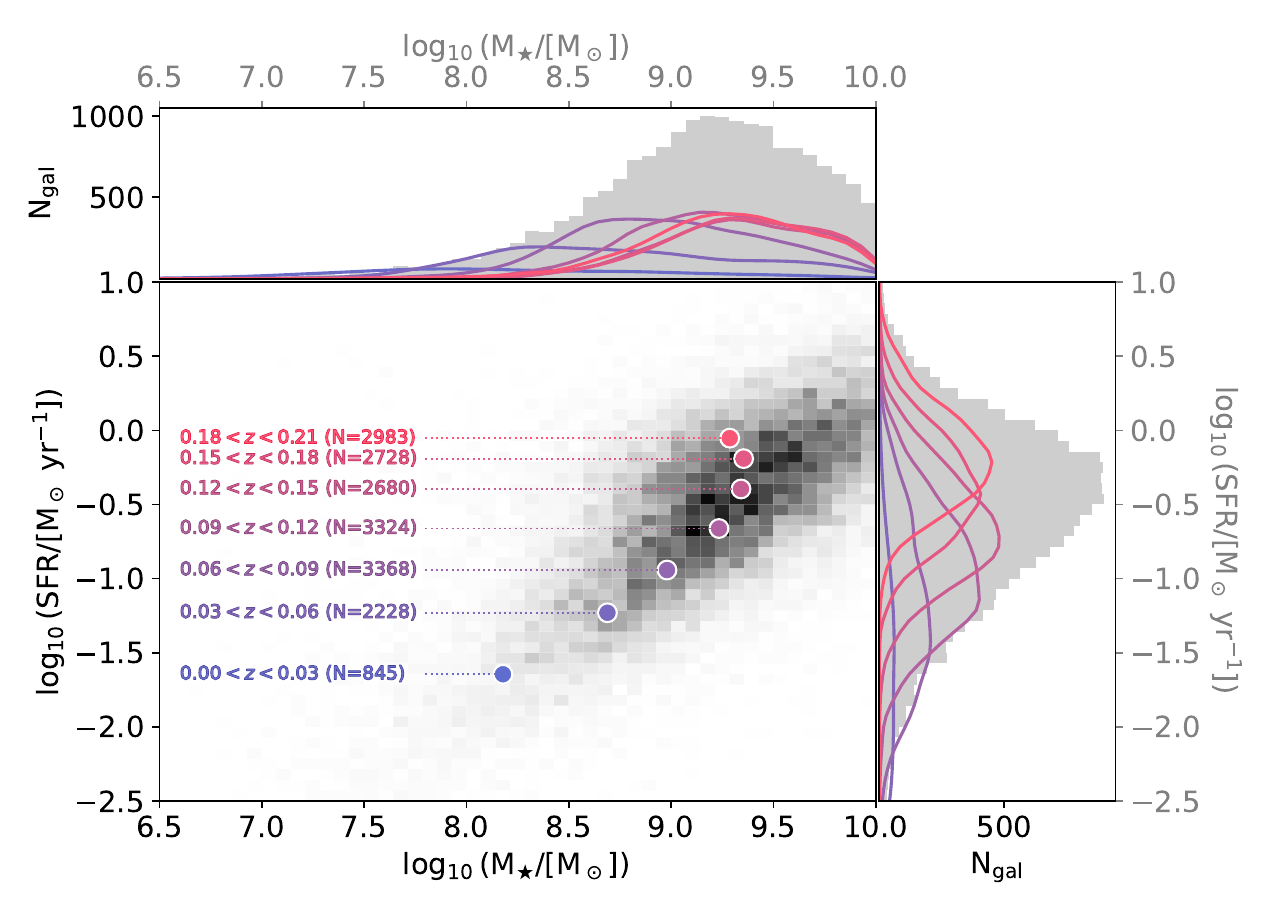}
  \caption{ 
      The distribution of the \sagabg{} sample as a function of 
      stellar mass and \halpha{}-derived star formation rate. 
      The clear evolution of the sample in $M_\star$-SFR space is a combination of
      physical evolution of the galaxy population with redshift and redshift-dependent 
      selection effects.
      We show the
      joint distribution in the main panel and projected distribution 
      over stellar mass and SFR at top and right, respectively, in grey. 
      In the main
      panel, we additionally show the median value of the sample as a 
      function of redshift; Gaussian kernel density estimates of these 
      projected distributions are shown at top and at right with the same 
      colors. We note that the normalization of the kernel density estimates are chosen to maximize the visibility of the density estimate, and are not reflective of the number of galaxies in each redshift bin.
      }\label{f:sampledescription}
\end{figure*}

\autoref{f:sampledescription} shows the distribution of the sample as a function of
stellar mass and \halpha{}-based star formation rate. We 
show the median values of both \logmstar[] and $\log_{10}{\rm SFR/(M_\odot\ yr^{-1})}$ as a function of redshift, as indicated in the main panel. As expected, the median stellar mass and star formation rate of the 
sample increases with increasing redshift; the crux of this paper is to 
determine how much, if any, of the change in the sample in $\rm M_\star-$SFR space can be attributed to a physical 
evolution in the SFS.

We make a more permissive cut on the precision at which lines must be detected (and the necessity of detecting weak lines) on the \sagabg{} sample in this work as compared to \citetalias{kadofong2024}. Because we are concerned only with modeling the \rrr{population-level} evolution of the SFS and not with the chemical evolution of the sample, we do not require \rrr{all individual} galaxies in the present sample to have a well-constrained estimate of internal reddening. \rrr{However, we note that 70\% of the galaxies in the sample do have a measured ($\sigma_{A_{\rm H\alpha}}<1$) optical extinction.}

\rrr{For the galaxies without a well-measured optical extinction}, we require only that $\sigma_{A_{H\alpha}}< 5$. However, for galaxies with $\sigma_{A_{H\alpha}}\geq 1$
we adopt a constant value of $A_{H\alpha} = 0.43$, which is the median value of $A_{H\alpha}$ \rrr{(with standard deviation $\sigma_{A(H\alpha)}=0.33$)} for the galaxies with well-constrained internal extinction and consistent with the adopted value for the SAGA satellites \citep[see Equation 4 of ][]{sagaiv}. \rrr{We note that instead excluding galaxies without well-measured optical extinctions would not make a statistically significantly impact on the results of this study.}

We do not see strong evidence that would advocate for the adoption of a stellar-mass- \rrr{or sSFR-}dependent form of $A_{H\alpha}$ over the stellar mass range considered. We find that \rrr{using this} modified sample \rrr{with the SFS-only model we will describe below} results in a slightly shallower evolution of the SFS \rrr{as compared to \citetalias{kadofong2024}}, as we will present below, but \rrr{that running the \citetalias{kadofong2024} model with the modified sample} does not affect the estimate of the mass-loading factor (the main result of \citetalias{kadofong2024}).

This more lenient cut on the \sagabg{} sample produces a final sample of \nfinal{} galaxies, representing an increase of 6368 galaxies over the sample presented in \citetalias{kadofong2024}.

\section{Modeling the SFS}\label{s:modeling}
We take a differential approach to modeling the observed evolution of our sample in order to disentangle the physical evolution of the SFS from distance-based observational effects.
In \citetalias{kadofong2024} we considered a broader version of this framework to constrain the mass-loading factor of low-mass galaxies by tracking their chemical evolution; in this work we will concern ourselves only with the change in the sample in the SFR-$M_\star$ plane.


Instead of attempting to assemble a population of realistic galaxies, we will use the $z\sim 0 $ galaxies in \sagabg{} and our knowledge of the \sagabg{} selection function to ``retrogress'' our lowest redshift galaxies to larger distances by modeling the differential effects that act on galaxies as a function of lookback time. These effects include both physical changes in galaxy properties (here the SFR and stellar mass, but in \citetalias{kadofong2024} we also consider gas-phase oxygen abundance) and 
observational effects (e.g., changes in apparent magnitudes due both to 
increased distance and $k$-corrections). 

This approach makes the following assumptions:
\begin{description}
  \item [Relative completeness] We assume that our reference sample of galaxies at $z<0.035$ is the most complete subset of our sample. 
  \item [Evolutionary link] We assume that evolving the reference sample to higher lookback time should reflect the higher redshift samples once observational effects are taken into account. That is, we assume that there exists an evolutionary link between the reference sample and higher redshift bins. We will expand upon the exact form of the evolution in the following section.
  \item [Accretion] We assume that the accretion of stellar mass from infalling satellite systems is negligible compared to in-situ star formation within the last \tlbmax{}.
\end{description}

These assumptions in our model are the same as \citetalias{kadofong2024}, though we note that some assumptions 
presented in \citetalias{kadofong2024} are not necessary for this work (in particular, the assumptions about the rate at which the gas reservoir of low-mass galaxies is replenished by cosmic inflow). 

The assumption of relative completeness is straightforward to justify given the photometric selection of the \sagabg{} sample; more distant galaxies will have lower \halpha{} fluxes for a given \halpha{} luminosity, and the 
observational reddening in the $g-$ and $r-$band will cause us to 
select intrinsically bluer and more star-forming galaxies at fixed stellar mass and increasing redshift. 

The second assumption, that the galaxy samples at different redshifts are  connected in an evolutionary sense, is justifiable only at the statistical level --- the assumption does not hold true for individual galaxies. 
Of particular interest for the evolution of the SFS over time, self-quenching (that is, quenching due to internal processes alone) is a relatively small effect in the stellar mass range of interest \citep{geha2012}. 
\rrr{The presence of a massive host is well-demonstrated to drive quenching \citep[see, e.g.,][]{wang2018, sagaiv}, but based off the analysis of \citetalias{kadofong2024}, we exepct a relatively small and redshift-independent fraction ($<10\%$) of the \sagabg{} galaxies to be satellites. We therefore do not expect the contribution of redshift-dependent host-satellite 
interactions to cause our sample SFS to significantly deviate from this assumption.}

Finally, due to a declining stellar mass-to-halo mass relation, 
low-mass galaxies are not generally expected to accrete significant stellar mass via minor mergers \citep[e.g.][]{purcell2007, conroywechsler2009, brook2014}. 
Though major mergers between dwarf galaxies do occur, these are relatively rare at low redshift, with
pair and merger searches yielding an expected merger incidence of a few percent \citep{stierwalt2015,besla2018, kadofong2020a}. We thus expect stellar mass build-up due to accretion to be 
small with respect to in-situ star formation 
in the past \tlbmax{} for our sample on a statistical level.

\subsection{Sample retrogression and comparison}\label{s:modeling:retrogression}

To model the change of the SFS as a function of lookback time, we must make an assumption about the form of its evolution. 
We assume
that the average star formation history of the galaxy sample can be approximated as a linear enhancement
in the past \tlbmax{}. In detail, this means that we may write the star formation rate of a given reference sample galaxy 
in terms of its $z<0.035$ SFR and $\asfh{}$:

\begin{equation}\label{e:sfh}
  \text{SFR}(t) = \text{SFR}(t=0)[1+\asfh{}t],
\end{equation}
where $t$ is the lookback time \rrr{in Myr} (i.e, $t=0$ is the present day)\rrr{, and $\asfh{}$ is the factor by which the SFR changes per Myr for the average galaxy (i.e., the bulk redshift dependence of the SFS). One can also note that $\asfh{}$ is the inverse of the characteristic timescale $\tau$ for a exponentially declining SFH in the linear approximation}. 
This very simple form of the SFS evolution is chosen for two reasons: first, with some degree of hindsight, we find that the evolution of the SFS is well-fit by \autoref{e:sfh}.  A more complex form is therefore not well-motivated by the data at hand. Second, because we are using real galaxies as the boundary condition for this problem, the scatter in the SFS should already be present in the modeled sample. We thus eschew a form of the SFS evolution that includes a stochastic component. 

We place a uniform prior on $\asfh{}$ of $\text{Pr}[\asfh{}] = \mathcal{U}(-4,4)$ Myr$^{-1}$; our choice of prior bounds is selected to be physically reasonable and does not affect our results. \rrr{This prior, which includes $\asfh{}<0$, allows for bulk decreasing SFRs as a function of redshift. Mathematically, this could lead to negative, non-physical star formation rates at sufficiently low $\asfh{}$ and high redshift. In practice, the data strongly exclude such a scenario.}

Given the star formation rate of a galaxy at $t=t_i$, we then predict how the stellar mass has changed from lookback time $t_i$ to a more distant lookback time $t_{i+1}$.  The change in stellar mass can then be written as:
\begin{equation}\label{e:mstartlb}
  \dot M_\star(t_i) = \text{SFR}(t_i).
\end{equation}
Note that because we are going backward in cosmic time (increasing lookback time), $\Delta t<0$ and so $\dot M_\star \Delta t < 0$.

We divide our sample into bins out to $z=0.21$ 
(\tlb{}$=2.5$ Gyr). Our reference sample is comprised of the galaxies in the first redshift bin, at $z<0.035$, equivalent to $\text{\tlb}<480$ Myr. 
\rrr{We defined this redshift limit for the reference sample in \citetalias{kadofong2024}; we will revisit whether this definition of the reference sample 
lies within the $z\sim0$ approximation in \autoref{s:discussion:zsimzero}. }
This first bin contains \nreference{} galaxies. We apply \autoref{e:sfh} and \autoref{e:mstartlb} at each lookback time in order to estimate the expected underlying SFS. 

As in \citetalias{kadofong2024}, we apply the effect of our 
selection function to evaluate whether our retrogressed galaxies would be 
observable (that is, whether the SAGA survey would have succeeded in obtaining a secure redshift) at a given lookback time. We develop our empirical 
estimate of the selection
function in Section 4.1 of \citetalias{kadofong2024};
as we have done throughout the manuscript, we will summarize the salient points of the selection function here but refer the reader to \citetalias{kadofong2024} for additional details.

The SAGA spectroscopic targeting scheme is primarily based upon the apparent $r$-band magnitude, $r$-band surface brightness, and $g-r$ color of the target. In \citetalias{kadofong2024}, we determined the probability that a potential target would be targeted in the \sagabg{} sample as a function of its photometric properties. The expected photometric properties as a function of redshift are then calculated for each reference, and the galaxy is stochastically \rrr{``}targeted\rrr{''} with the prescribed targeting probability.

In brief detail: we compute $m_r(z)$, $m_g(z)$, and $\mu_{r,\rm eff}(z)$ via $k$-corrections
estimated directly from the spectra. \rrr{As discussed in Section 4.4 of \citetalias{kadofong2024}, we do not include the effect of stellar mass or age given that these effects are expected to be small over the redshift range considered.}
In our estimation of $\mu_{r,\rm eff}(z)$, we assume that the physical size of the galaxy remains the same as a function of redshift. We also tested a model in which the galaxies were constrained
to lie on the $z=0$ stellar mass--size relation of \cite{carlsten2021}. The 68\% confidence intervals of the parameters inferred with this version of the model were consistent with our fiducial model but provided a significantly worse fit to the observed samples (as implied from the likelihood of the posterior sample) as compared to our fiducial model. 

For each spectrum, we additionally estimate the \halpha{} detection limit;
to be flagged as observable, we require that the retrogressed galaxy be targeted and that its \halpha{} flux \rrr{as predicted from its predicted SFR} at the retrogressed redshift exceeds the 
$3\sigma$ detection limit defined \rrr{in \autoref{s:methods:emlines}}.  
\rrr{These criteria allow us to construct the probability that any given retrogressed galaxy will be photometrically selected as a SAGA survey target, and will achieve redshift success upon observation given the data quality of the SAGA survey. These successfully ``targeted'' galaxies make up what we will call our observable retrogressed model predictions; i.e., our backwards-integrated model as we would expect to view it through the observational lens of the SAGA survey. We will now construct the likelihood of observing the real \sagabg{} sample by constructing a kernel density estimate of this retrogressed model prediction. That is, we compare the observable retrogressed model predictions to the real \sagabg{} sample as a function of redshift.}

Following the notation of \citetalias{kadofong2024}, we write 
the likelihood \rrr{of the observed \sagabg{} sample given the physical evolution of the SFS} as:

\begin{equation}
    \ln\mathcal{L}_j(\vec{\text{SFR}},\vec M_\star|\asfh{},z_j) = \sum_i \ln\mathcal{L}_{j}^{{\rm SM}}(\text{SFR}_i,M_{\star,i}|a_{\rm sfh}, z_j)
\end{equation}
where the $i$ subscript indicates the $i^\text{th}$ observation in a given 
redshift bin, and $z_j$ is the $j^\text{th}$ redshift bin.
$\mathcal{L}_j^{\rm SM}$ is the likelihood of the data in SFR-stellar mass space.
We \rrr{adopt a non-parametric} estimate \rrr{of $\mathcal{L}_j^{\rm SM}$} 
\rrr{by constructing the} Gaussian kernel density estimate of 
our observable retrogressed model predictions with bandwidth $N^{-1/(d+4)}$ (i.e., Scott's Rule, \citealt{scott1992}, where $d=2$ is the dimensionality of the sample). \rrr{This formulation allows us to} then compute the likelihood of observing the real \sagabg{} sample at each redshift bin \rrr{as the probability density of the gKDE at the location of the observed \sagabg{} galaxy}.

\begin{figure*}[t]
  \centering     
  \includegraphics[width=\linewidth]{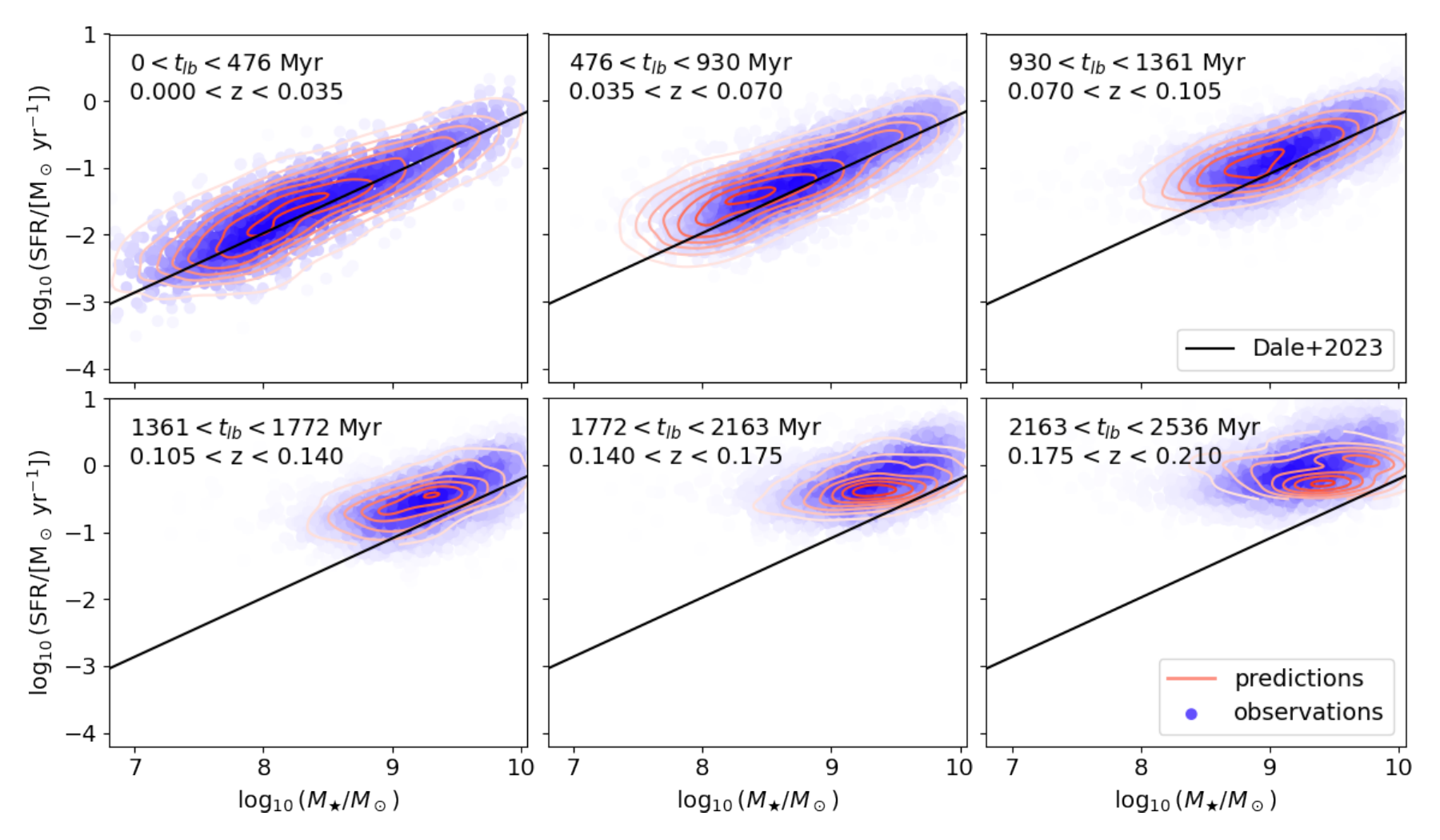}
  \caption{ 
      A comparison between the \rrr{observable retrogressed model predictions} (\rrr{the expected distribution of our median posterior model estimate as would be observable after the SAGA survey selection function}, red contours) and observed SAGA background sample \rrr{(purple scatter)}
      as a function of redshift in the stellar mass-star formation rate plane. 
      Here we also show the SFS as measured from the Local Volume Legacy
      (LVL) Survey by \cite{dale2023} (black line) as a comparison to the $z=0$ SFS. 
      }\label{f:SFS_evolution}
\end{figure*}

We use the \textsf{emcee} implementation of the Affine Invariant Markov Chain Monte Carlo ensemble sampler \citep{emcee} for parameter inference. We run the chain for 10000 steps with 32 walkers and visually confirm chain convergence.

Our model of the selection function for the \sagabg{} galaxies is similar to the 
incompleteness model developed in \cite{sagaiii} for the main SAGA science sample
(that is, the satellites of the 101 SAGA hosts). Although both 
selection functions are based on the same spectroscopic dataset, the models differ in their definitions of the selected samples. 

\subsection{Fitting the SFS}\label{s:methods:fitting}
 We fit our inferred star-forming sequences with a power-law of the form:
\begin{equation}\label{e:plaw}
  \log_{10}\left(\frac{\text{SFR}}{[M_\odot\text{ yr}^{-1}]}\right) = \alpha\left(\log_{10}(\frac{M_\star}{[M_\odot]}) -8.5 \right)+\text{SFR}_{8.5}
\end{equation}
where 
$\alpha$ is the power-law index and SFR$_{8.5}$ is the star formation rate at \logmstar[$=8.5$]. 

In each redshift bin, we adopt a likelihood that assumes that both the measurement error and underlying physical model are well-described by Gaussians at fixed stellar mass; that is:
\begin{equation}
  {\rm \color{red}{\mathcal{L}}}[{\rm SFR}_i|m_i, \sigma, s_i] = \frac{\rm Exp\left[{\frac{-({\rm SFR}_i-m_i)^2}{2(s_i^2+\sigma^2)}}\right]}{\sqrt{2\pi (s_i^2+\sigma^2)}},
\end{equation}\label{e:likelihood}
where $m_i$ and $\sigma$ describe the mean $\log_{10}({\rm SFR}/[M_\odot\ {\rm yr}^{-1}])$ and its intrinsic scatter at the stellar mass of the $i^{\rm th}$ galaxy in the retrogressed reference sample, and $s_i$ is the uncertainty on its star formation rate ${\rm SFR}_i$ as determined from the uncertainty on the \halpha{} luminosity. This uncertainty is estimated as the difference between the 84\textsuperscript{th} and 16\textsuperscript{th} percentiles of the line fit.

{}

We again use the \textsf{emcee} implementation of the Markov Chain Monte Carlo sampling algorithm \citep{emcee} to sample the posterior given a uniform prior over all fit parameters where $\sigma$ is required to be positive. We run the chains for 500 steps with 32 walkers and discard the first 100 steps of each walker to arrive at 
our parameter estimates; as before, we visually confirm chain convergence.

\section{Results}\label{s:results}
Having laid the groundwork of our model, we turn now to its
application to the sample at hand. Because this work only concerns the 
redshift evolution in the SFS, the reader interested in the change in 
gas-phase metallicity of the \sagabg{} galaxies should refer to \citetalias{kadofong2024}.

\subsection{The Observed SFS}\label{s:results:SFS}
Here we discuss the observed evolution of the SFS in both our observations and \rrr{observable retrogressed model predictions, as defined in \autoref{s:modeling:retrogression}}. 
The observed evolution is a product of both the physical evolution and the effect of our observational limitations and selections on the $M_\star-$SFR plane as a function of redshift.

Each panel in \autoref{f:SFS_evolution} shows a redshift slice of $\Delta z=0.035$ ($370<|\Delta t|<480$ Myr) with our observed sample shown by the blue points (colored by a Gaussian kernel density estimate to visually indicate density) and our model predictions shown by red contours. In each panel, we also show \rrr{measurements of the Nearby Universe} SFS \rrr{(which, we remind the reader, we use to refer to galaxies within tens of Mpc from the MW)} as measured for the Local Volume Legacy Survey (LVL) by \cite{dale2023}. There is a 
small but statistically significant offset \rrr{at $z<0.035$} with respect to the LVL SFS
that we will return to in the following section. \rrr{In \autoref{s:appendix:modelcomparison}, we} show the \rrr{difference between the} median of the observed sample \rrr{and the observable retrogressed model prediction} \rrr{for our fiducial model, as well as a model that assumes and exponentially declining SFR$(t)$ and a constant average SFR}. We \rrr{find that the exponentially declining $\tau$ model prefers a $\tau$ that renders the model results statisically indistinguishable from our fiducial model. We also find that all three models}
slightly underestimate the average stellar mass and 
star formation rate of the sample at $0.035<z<0.070$, but otherwise find 
agreement between our \rrr{fiducial} model and \rrr{the} observed sample.

The concordance of the observations and model predictions supports our choice of a simple approximation to the mean recent SFH of the galaxies in our sample (see \autoref{e:sfh}). As a reminder, the change of the SFS shown in \autoref{f:SFS_evolution} is a combination of the physical evolution in the SFS and the redshift-dependent observational selection function imposed by our survey completeness. 
In \autoref{s:discussion:SFS}, we will examine the implications of the physical component of this
shift.

\subsection{The Reference SFS at $z<0.035$}\label{s:results:localSFS}
In this analysis, we compare the evolution of the SFS to \rrr{our lowest-redshift measurement of the SFS at $z<0.035$, thus assuming that our \sagabg{} SFS at $z<0.035$ is equivalent to a $z\sim 0$ SFS}. However, it is important to note that the $z<0.035$ sample
is itself subject to observational effects that may affect our estimate of the SFS.
We now compare our best-fit SFS to SFR-$M_\star$ relationships estimated from 
samples of galaxies in the Nearby Universe to understand whether incompleteness drives a significant bias in the \sagabg{} $z\sim 0$ SFS and to test the assumption that $z\sim 0$ is 
comparable to $z<0.035$ given the statistical precision of our sample.


In \autoref{f:localSFScomparison}, we compare to three Nearby galaxy 
samples: the $z = 0$ Multiwavelength Galaxy Synthesis (Z0MGs) sample \citep[$d\lesssim 50$ Mpc, $z\lesssim 0.01$;][]{leroy2019}, the Local Volume Legacy (LVL) survey \citep[$d<11$ Mpc, $z\lesssim 0.0025$;][]{dale2023}, and the SAGA satellites themselves 
\citep[$d\lesssim 40$ Mpc, $z\lesssim 0.009$;][]{sagaiv}.
Both Z0MGs and LVL star formation rates are derived from a UV-to-IR SFR prescription, while the star formation rates of the SAGA satellites are
measured from NUV luminosities.

There is generally good agreement between our estimate of the low-redshift SFS (solid blue line) and those of~\cite{leroy2019}  (grey solid line)
and~\cite{dale2023} (red solid line). At low stellar masses (\logmstar[$<9$]), our 
estimate of the SFS is significantly lower than an extrapolation of \cite{leroy2019}, but the majority of the \cite{leroy2019} sample is at higher stellar masses relative to \sagabg{} and the \cite{dale2023} sample, as shown in the upper panel of \autoref{f:localSFScomparison}. 
Though the bottom panel of the same figure shows that there are differences between each SFS, 
these are $\lesssim 0.2$ dex across the stellar mass range covered, which is
small compared to the \rrr{paper-to-paper} dispersion in observed \rrr{fits to the} SFS (see, for
example, \autoref{f:SFS_datalfits}). \rrr{The dispersion in the absolute calibration of the SFS seen in the literature, which is likely due to both systematic uncertainty in the SFR and stellar mass calibrations, underscores the utility of measuring the SFS redshift evolution using the same stellar mass and SFR measurement methodology, especially in the low-$z$ regime where the magnitude of the redshift evolution may be similar to the magnitude of these calibration uncertainties.}
We additionally find that our low-redshift SFS is within $\sim 0.1$ dex of the SFS measured for the SAGA satellites by \paperiv{} using NUV luminosities (dotted blue line). The SFS is in similar agreement with the satellite SFS measured from H$\alpha$ equivalent width, which we do not show in \autoref{f:localSFScomparison}. 

\subsection{The Underlying SFS}\label{s:results:underlyingsfs}

In \autoref{s:results:SFS} (and \autoref{f:SFS_evolution}), we showed that our \rrr{observed retorgressed model predictions are} 
successful in reproducing the observed change in the 
SFS as probed by the \sagabg{} sample, \rrr{and in \autoref{s:results:SFS} (and \autoref{s:results:localSFS}), we showed} that our $z\sim 0$ SFS is unbiased with respect to highly complete local samples of star-forming galaxies. Having demonstrated this, we now use our model predictions \textit{without} applying a \sagabg{}-like observability cut to isolate the physical component of the observed change in the sample in $\rm M_\star$-SFR space. \rrr{We will refer to these model predictions, which should reflect the underlying relation between stellar mass and star formation rate that produces the \sagabg{} sample, as the \sagabg{} SFS.}

We fit the SFS at each redshift bin in the same way as we did for the $z<0.035$ sample using \autoref{e:plaw}. The results of this exercise are shown in \autoref{f:SFS_datalfits}. Here we show in blue the SFS \rrr{inferred from the} \sagabg{} sample in bins of increasing redshift from top left to bottom right. The shaded regions in each panel show the best-fit intrinsic scatter of the \sagabg{} SFS ($\sigma_{\rm SFR}$). The median posterior values
(and 16\textsuperscript{th}-to-84\textsuperscript{th} percentile ranges) inferred for each parameter are given in the bottom right of each panel. These fits are a simple power-law of the same form as \autoref{e:plaw}. We 
show only the range in the stellar mass between the $5^{\rm th}$ and $95^{\rm th}$ percentile of the observed \sagabg{} sample in each panel. For the reader's convenience, we also show the $0<z<0.035$ 
\sagabg{} best-fit SFS in each panel as a thin \rrr{dotted} line. \rrr{We give the best-fit parameters for each redshift bin in \autoref{s:appendix:fits}.}

Two main points are apparent. First, the \sagabg{} SFS is consistent with literature results at the level of the existing study-to-study scatter in SFS determinations. This, along with our comparison to local SFS measurements in the preceding section, indicates that our measurement of the SFS is comparable in star formation rate completeness as previous studies in this regime.  Second, there is significant redshift evolution (a redshift-independent SFS is excluded at the 95\% confidence level) to higher mean SFRs in the \sagabg{} SFS from $z\sim 0$ to $z\sim 0.2$. This redshift evolution is echoed in the observational literature; its implications for both Nearby Universe studies and dwarf assembly will be examined in the following sections.

We fit the redshift evolution of SFR$_0$, $\alpha$, and the intrinsic scatter $\sigma_{\rm SFR}$ with a linear model 
employing the same inference framework as 
presented in \autoref{s:methods:fitting}. We find evidence for an increasing normalization of the SFS, SFR$_{8.5}$, at the $>\!90\%$ confidence level. There is weak evidence for an increase in the intrinsic scatter of the SFS as a function of redshift, but the 90\% confidence level includes the 
constant-scatter case. We see no evidence for a change in the power-law index of the SFS over our redshift range. In particular, we find:
\begin{equation}\label{e:parameterevolution}
    \begin{split}
        \alpha(z) &= -0.13^{+0.32}_{-0.30}z + 0.80^{+0.03}_{-0.04}\\
        {\rm SFR}_{8.5}(z)&=1.24^{+0.25}_{-0.23}\ {\rm z} -1.47^{+0.03}_{-0.03}\\
        \sigma_{\rm SFR}(z)&=0.22^{+0.16}_{-0.16}\ {\rm z} + 0.38^{+0.02}_{-0.02},
    \end{split}
\end{equation}
where the reported uncertainties reflect the 16\textsuperscript{th} to 84\textsuperscript{th} percentile range of the posterior. We thus find evidence for a significant change in the normalization of the SFS, but conclude that the sample does not produce compelling evidence to argue for a changing SFS power-law index or scatter over the redshift range considered.

\section{Discussion}
\begin{figure}[t!]
  \centering     
  \includegraphics[width=\linewidth]{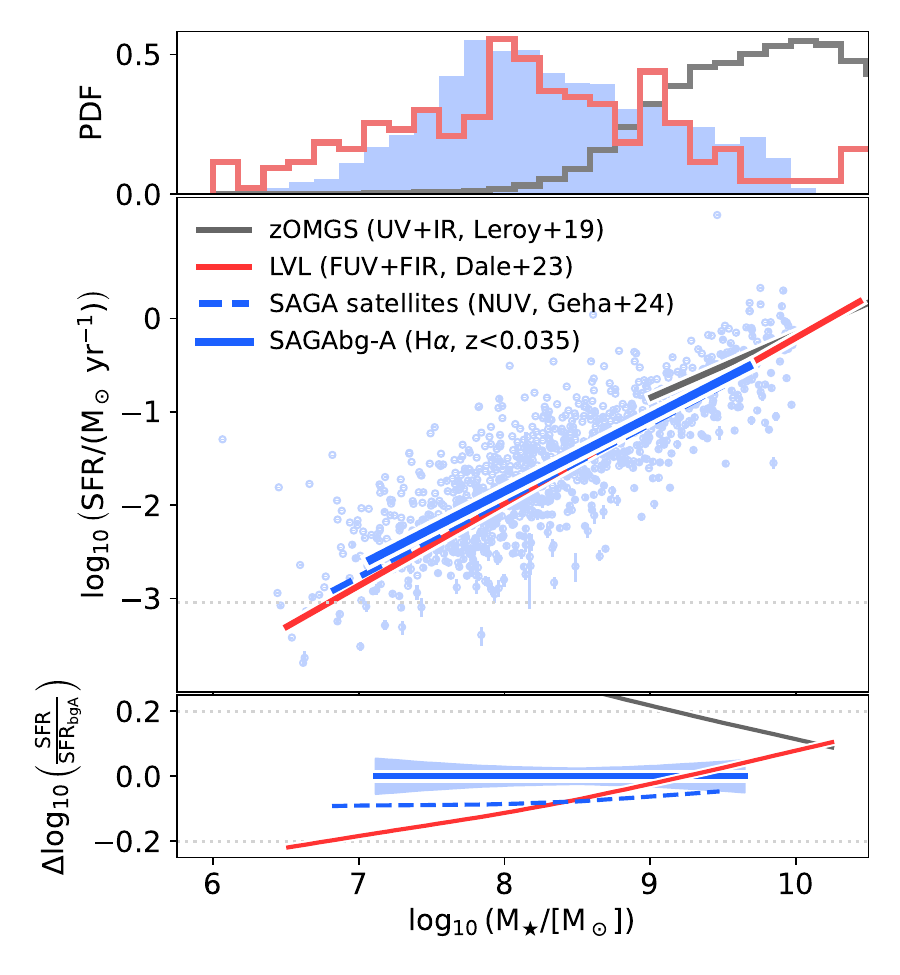}
  \caption{ 
      A comparison between the SFS that we estimate for our reference sample ($z<0.035$, blue solid line) the local star-forming sequences of the Local Volume Legacy Survey \citep[red,][]{dale2023}, the $z=0$ Multiwavelength Galaxy Survey \citep[grey,][]{leroy2019}, and the SAGA satellites (dashed blue line, \citet{sagaiv}). 
      The upper panel shows the distribution over the stellar mass of each sample.
      The main panel shows the best-fit SFS of each sample, as well as individual 
      points of the \sagabg{} sample at $z<0.035$.
      The horizontal line shows our approximate star formation rate
      detection limit at $z=0.035$. Each SFS is shown for the central 90\% 
      stellar mass range of its respective sample in the main panel. The bottom 
      panel shows the logarithm of the ratio between each comparison SFS and the 
      SFS derived from \sagabg{}. 
      }\label{f:localSFScomparison}
\end{figure}

\begin{figure*}[t]
  \centering     
  \includegraphics[width=\linewidth]{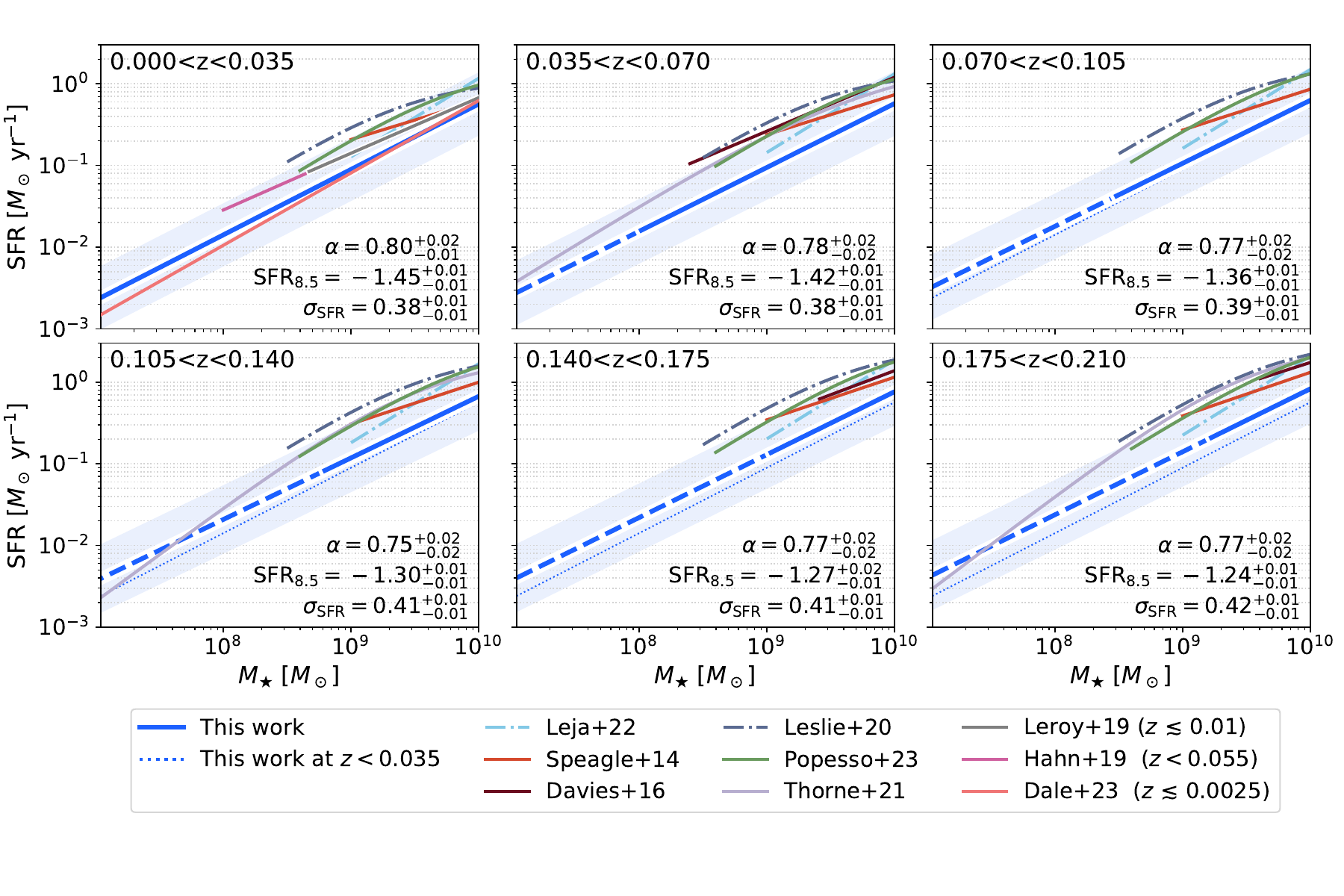}
  \vspace{-30pt}
  \caption{ 
      A comparison between our model SFS (blue) and other results from the observational literature at matched redshift. Our best-fit SFS is shown as a dashed curve where the average retrogressed galaxy is expected to be $m_r>21$, the approximate spectroscopic targeting limit of SAGA.
      We show the best-fit intrinsic scatter of the \sagabg{} SFS as a blue-shaded region; the parameter 
      estimates of \autoref{e:plaw} are given for the \sagabg{} SFS in each panel. As a visual reference, we also show the $z<0.035$ SFS fit for the \sagabg{} in each panel as the dotted blue line. 
      We also show fits from the observational
      literature \citep{speagle2014, davies2016, hahn2019, leroy2019, leslie2020, leja2022, dale2023}. Extrapolated fits to SFS fits made at higher
      redshifts are shown by dot-dashed curves, and SFS fits to only the $z\approx 0$ Universe are shown as dashed curves. All other SFS comparisons 
      are shown as solid curves.
      }\label{f:SFS_datalfits}
\end{figure*}

Although it is common practice to bin galaxies from $0\leq z \lesssim0.2$ in low-redshift studies, the difference in lookback time between $z=0$ and $z=0.21$ is comparable to the difference in lookback time between $z=1$ and $z=2$. As such, understanding the evolution of the SFS at low-redshift
is important both for understanding metrics that characterize the star formation cycle, to constrain our models of galaxy evolution, and to understand the physical meaning of comparisons between low-redshift and Nearby Universe galaxy samples. These motives are especially important for low-mass galaxies given that our view of the galaxy population at \logmstar[$\lesssim 9$] has only recently begun to systematically expand past $z\sim 0$ \citep{pasquet2019, darraghford2022, dey2022, luo2023}. 

\subsection{Defining the $z\sim 0$ approximation}\label{s:discussion:zsimzero}

In \autoref{s:results:underlyingsfs}, we presented evidence for a measurable evolution in the SFS of low-mass galaxies from $0<z<0.21$, given the measurement precision enabled by the \sagabg{} sample. The existence of evolution 
is not in and of itself a surprising result; indeed, it is entirely expected that the galaxy population would
continue to evolve at $z\sim 0$. Our result does, however, have a marked implication for the use of the term ``$z \sim 0$'' itself. 

As the precision to which we can measure galaxy properties at
$z>0$ increases, so does the need to understand the contribution of redshift evolution to our comparisons between Nearby and low-redshift samples. Accounting for redshift evolution is especially important for studies that aim to disentangle the impact of our local cosmic geography on galaxy evolution --- that is, the effect of the presence of (e.g.) the Local Group, the Local Void, or Virgo Supercluster --- by way of comparison to low-redshift galaxy samples that are able to average over cosmic variance due to the large volume coverage.

The redshift at which we may say that the galaxy population has evolved 
significantly is clearly dependent on the precision at which we are measuring that 
evolution, and is thus dependent on the sample measurements themselves. Thus, ``$z\sim0$'' is a statement that depends not only on the physics of the system, 
but on the sample that is used to probe that physics. 

 In our analysis, we have made the assumption that galaxies at $z\leq 0.035$ are, 
at the level of precision attainable with the \sagabg{} sample, statistically 
indistinguishable in average SFR with respect to galaxies at $z=0$. 
\autoref{f:localSFScomparison} tests this assumption by comparing our 
measured $z\sim 0$ SFS with existing literature SFS measured for more nearby samples.
The agreement with our estimate of the SFS indicates that our reference sample does not incur a significant bias in $M_\star$--SFR space relative to samples that are complete down to lower stellar masses and star formation rates. This point is especially important because although we explicitly correct for incompleteness in our higher redshift samples, our completeness corrections are always relative to our most-complete reference sample. A biased estimate of the SFS at $z\sim 0$ would therefore propagate to a bias in the model estimate of the underlying SFS at higher redshift.
Consistency between our $z<0.035$ sample and Nearby Universe samples indicates that it is reasonable to compare the inferred evolution of the SFS as traced by our retrogression model to literature measurements of SFS evolution.

We can also measure the maximum redshift at which 
cosmic evolution is \textit{not} statistically significant in the \sagabg{} SFS. To estimate this redshift, we draw from the posterior distribution of our fit to the redshift evolution of the SFS parameters (as described in \autoref{e:parameterevolution})  and measure the evolution in SFS normalization as $\Delta {\rm SFR}_{8.5}(z) \equiv {\rm SFR}_0(z) - {\rm SFR}_0(0) $ where  ${\rm SFR_0}(z)$ and ${\rm SFR_0}(0)$ are sampled independently from the
posterior. We find that ${\rm SFR_0} (z)=0$ is ruled out with 95\% confidence 
at $z=0.049$. We thus define the redshift limit of $z\sim 0$ for the \sagabg{} sample to be $z=0.05$.

This is both a statement about the rate at which the SFS evolves over
cosmic time and the precision to which we can measure that evolution with the
\sagabg{} sample. Nevertheless, this approach provides a simple metric to evaluate whether scaling relations (including but not limited to the SFS) measured from low-redshift samples such as \sagabg{} 
can be used to contextualize Nearby Universe samples without confounding 
redshift evolution with other processes governing galaxy evolution such as environment \citep{sagaiv} or star formation feedback \citep{kadofong2024}. 

{}

\begin{figure*}[t]
  \centering     
  \includegraphics[width=0.95\linewidth]{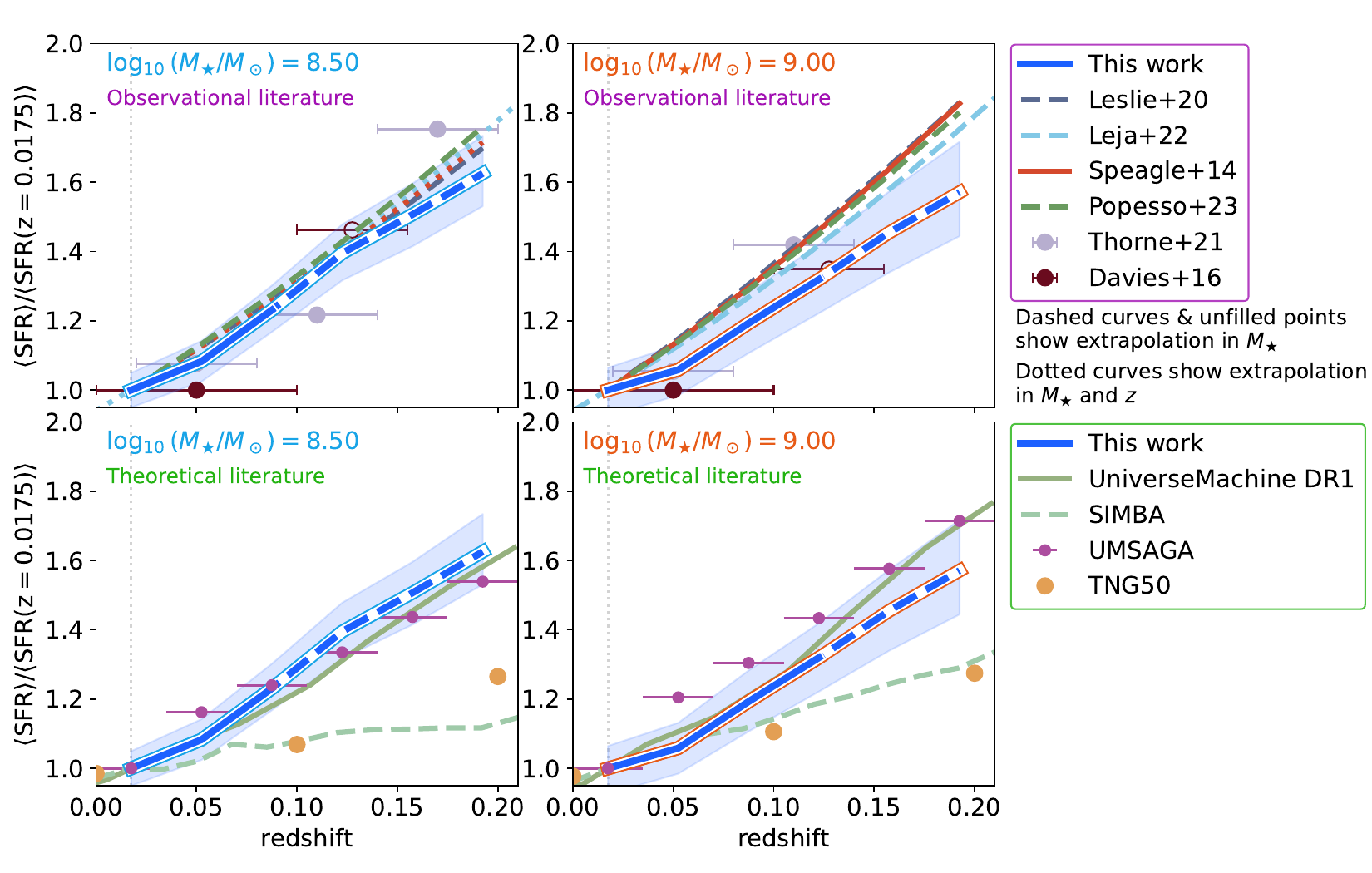}
  \caption{ 
      In each panel, we show the redshift evolution of the mean star formation
      rate at fixed stellar mass as inferred from \sagabg{}. The bold curve shows our 
      estimate of the mean SFR of a galaxy of the specified stellar mass given our power-law fit to the SFS as a function of redshift;
      the curve is solid in the redshift range where the average retrogressed
      galaxy remains
      above the nominal magnitude limit of 
      the SAGA spectra, and becomes dashed for model predictions below this limit. 
      In the left column, we show the SFS evolution at \logmstar[=8.5] (bold curve with blue outlines); the right column shows the same for \logmstar[=9.5] (bold curve with orange outlines). The shaded
      regions around the bold curve in each panel show the 16\textsuperscript{th} to 84\textsuperscript{th} percentile confidence interval on our estimate of the mean SFR($M_\star$). To account for differences in the absolute normalization of the SFS, we 
      show the redshift evolution of the SFS relative to its mean value at $z=0.0175$, as shown
      by the dotted grey vertical line.
      We compare the results from the \sagabg{} sample to results from the observational literature \citep{speagle2014, leslie2020, thorne2021, leja2022,popesso2023} at top and the theoretical literature at bottom \citep{behroozi2019, dave2019, nelson2019, sagav}.
      Observational fits that we have extrapolated beyond the original work's sample 
      domain in stellar mass are shown as dashed curves; fits that are extrapolated 
      in both stellar mass and redshift are shown as dotted curves.
      }\label{f:sfratpivot_evolution}
\end{figure*}

\subsection{Observational Redshift Evolution at $0<z<0.21$}
A significant redshift evolution of the SFS between $z\sim0$ and $z\sim 0.2$ is not unexpected observationally; such an offset has been noted 
in previous comparisons between widefield surveys such as the GAMA Survey 
and local galaxy surveys such as the LVL
Survey \citep[Figure 10 in][]{cook2014}. However, 
these offsets have been difficult to interpret because they have 
largely manifested in comparisons between different surveys; 
given that the magnitude of the effect is small and that there are significant
systematic offsets in the absolute calibration of the SFS even at $z=0$, as discussed above, 
the degree to which such an effect should be attributed solely to systematic differences between surveys has been unclear.
\rrr{Indeed, the magnitude of the systematic uncertainties involved in comparing different SFR tracers implies that using the same SFR tracers (or a highly complete understanding of systematic offsets between SFR tracers) is likely necessary to measure the low-redshift evolution of the SFS.}

In \autoref{f:SFS_datalfits}, we compare the inferred SFS of \sagabg{} at different redshifts to various measured relations from the 
observational literature \citep{speagle2014, davies2016, leslie2020, thorne2021, leja2022,popesso2023}.
 For these samples we additionally make a cut at $\log_{10}({\rm SFR}/[M_\odot\ \rm yr^{-1}]) > -3.6$, which is equivalent to the median limiting \halpha{} luminosity in our sample at $z=0.0175$. We will return to a comparison to theoretical predictions in the subsequent section; at present, we will turn our attention to understanding the $z<0.21$ evolution of the SFS as expected from the observational literature.

The slope of the SFS is relatively consistent throughout the literature --- amongst the observational fits that we consider (including \citealt{leroy2019} and \citealt{dale2023}), we find a mean slope of the $z\sim0$ SFS of $\bar m = 0.79$ and a standard deviation of $\sigma_m = 0.17$. If we remove the studies with the minimum (\citealt{speagle2014}, $m=0.48$) and maximum (\citealt{leja2022}, $m=0.96$) slope estimates, the mean SFS slope rises slightly to $\bar m=0.83$ and the study-to-study standard deviation drops to $\sigma_m = 0.09$. Our estimate of the $z<0.035$ SFS yields a slope of $m = 0.80\pm 0.02$, consistent with the literature consensus given the systematic uncertainties incurred in its estimation.
\rrr{At higher stellar masses (\logmstar[$\gtrsim 10$]) there is evidence that the SFS is not well-fit by a single power-law; this flattening is visible in the high-mass ends of \cite{leslie2020}, \cite{thorne2021}, and \cite{popesso2023}. However, because we are interested in stellar masses significantly lower than the expected inflection point of the SFS \citep{popesso2019,daddi2022,popesso2023}, adopting a functional form that allows for a stellar mass-dependent flattening in the SFS does not produce a significantly different fit to the SFS relative to the single power-law fit assumption.}

As in the \sagabg{} sample, the 
majority of comparable works in the literature predict a significant redshift 
evolution between $z\sim 0$ and $z\sim 0.2$.  In order to better evaluate this evolution,
in \autoref{f:sfratpivot_evolution}, we show the mean star formation rate of 
\sagabg{} (bold blue curves) as compared to the same 
literature results as a function or redshift. In order to account for differences in SFR calibration and internal extinction corrections, we consider the redshift evolution normalized to the value of the SFS at the given stellar mass and $z = 0.0175$, which is the midpoint of our first redshift bin. 

Here, each column shows a different stellar mass at which the SFS is evaluated, at \logmstar[$=8.5$]{} and \logmstar[$=9.5$]{}. 
The top row of the figure, with which we will presently concern ourselves, shows a comparison to the observational literature. The bottom shows a comparison to the theoretical literature. The works shown in 
\autoref{f:sfratpivot_evolution} are the same as those shown in \autoref{f:SFS_datalfits}.

We find that the \sagabg{} results agree well with the majority of the observational literature. The shaded region around the blue curve denotes the 5$^{\rm th }$ to $95^{\rm th}$ percentiles of our estimate of the mean SFR at fixed 
mass --- not the intrinsic dispersion of the SFS at fixed stellar mass. 
The majority of observational works to which we compare are extrapolations in
stellar mass and/or redshift, as noted by linestyle in the legend in the second column from the right. 
We do note that for the massive end of our sample (\logmstar[]{$\gtrsim 9.5$}), the overlapping samples of \cite{speagle2014} and \cite{thorne2021} predict 
a steeper evolution of the SFS; this could be attributed to a SFR-dependent drop in
the targeting completeness of the \sagabg{} sample for massive galaxies. 

Our most direct comparison is to the work of \cite{davies2016}, who use the GAMA survey to estimate the evolution in the SFS for slightly higher stellar masses out to $z=0.35$. This work is shown by the orange errorbars (which indicate the redshift range over which the SFS is averaged) in \autoref{f:sfratpivot_evolution}; filled points indicate regions of stellar mass and redshift space where a typical galaxy (as defined by the color-mass relation used to calculate the \sagabg{} stellar masses) is brighter than the nominal GAMA survey depth of $m_r = 19.8$. Unfilled points indicate an extrapolation of the \cite{davies2016} SFS.

Thus, the concordance between the results of the \sagabg{} SFS and the observational 
literature indicates not only consistency, but also continuity in the 
behavior of the SFS as a function of both stellar mass and redshift down to 
\logmstar[]{}~$\sim 8.5$ and $z\sim 0$. 

\subsection{Comparisons to Theoretical Predictions}\label{s:discussion:SFS}
Having established that the redshift evolution that we see in the \sagabg{} sample is consistent with extrapolated expectations from the observational literature, we return 
to \autoref{f:sfratpivot_evolution} to consider a comparison to the theoretical literature. Here,
we compare to UniverseMachine DR1 in purple \citep{behroozi2019}, Illustris TNG50 in 
orange \citep{nelson2019}, and SIMBA in pink \citep{dave2019}.  We find that for the stellar mass
and redshift range covered, the updated UniverseMachine model that incorporates
observational constraints from the SAGA satellites \citep{sagav} produces a SFS that is consistent with the
SFS from UniverseMachine DR1.

The predictions from the two cosmological simulations considered diverge from our observational results at \logmstar[$\leq9$], as shown in the bottom panels of \autoref{f:sfratpivot_evolution}: we observe a significantly steeper evolution in the SFS than that which is predicted
by SIMBA and TNG50 up to $z\sim 0.2$. We stress that this comparison does not mean that 
the contemporary simulation landscape writ large is unable to reproduce the low-redshift
evolution of the SFS; rather, the difference between observational and theoretical results 
indicates that there does exist information in the $z<0.2$ low-mass galaxy population that
can help to constrain dwarf assembly over cosmic time.

The data-driven UniverseMachine models, meanwhile, predict redshift evolution that is broadly consistent with
our observational results  (within the 95\% confidence interval of our estimated median SFR in each redshift bin considered). UniverseMachine uses observational constraints and dark-matter-only simulations 
to estimate the form of the galaxy-halo connection in SFR-halo property space. We show both the \umdr{} results of \cite{behroozi2019} and the UMSAGA results of \cite{sagav}, who extended the UniverseMachine model to incorporate constraints from the main SAGA satellite survey. 
The agreement between \umdr{} and \sagabg{} is notable given that the only observational constraint used for \umdr{} that significantly overlaps with our sample mass and redshift range is a stacked 
SFS that spans the full redshift range of the sample \citep{behroozi2019}. 

A divergence between simulation-based and observed SFSs at low mass is perhaps unsurprising. First, the redshift evolution of the high-mass galaxy sample is generally better constrained than for low-mass galaxies because complete samples of star-forming galaxies extend to higher redshift with increasing stellar mass. The assembly of the high-mass galaxy population is thus better constrained at nearly all cosmic times relative to that of low-mass galaxies; possible exceptions are the very nearby Universe (due to the effect of a declining stellar mass function on a volume-limited sample) and the very early Universe (where samples tend to skew towards vigorously star-forming low-mass galaxies). Second, the effects of the star formation cycle are more pronounced in the structural and chemical evolution of low-mass galaxies, both because low-mass galaxies accrete relatively fewer stars from satellites 
and because it is thought that low-mass galaxies' shallower potential wells render them more sensitive to the effects of star formation feedback \citep{dekelsilk1986, elbadry2016, elbadry2018, kadofong2020b, kadofong2022a}. Subgrid models related to star formation physics may thus play a larger role in setting the redshift-dependent SFS at low stellar mass. 

\rrr{The redshift evolution of the low-mass SFS cannot uniquely identify the astrophysical cause for the divergence between the observed Universe and contemporary cosmological simulations, but these results do fit into the larger ecosystem dwarf galaxy observations that can elucidate the possible parameter space of physical mechanisms behind the current discrepancy.}
In \citetalias{kadofong2024} of this series, we found that the mass-loading factor ($\etam{}=\dot M_{\rm out}/\text{SFR}$) predicted by several current cosmological simulations is also more than an order of magnitude 
larger than the mass-loading factor we estimate for the \sagabg{} galaxies. 
A shallower evolution in the TNG50 SFS allows for a higher mass-loading factor while maintaining the same overall displacement of metals in outflows, allowing for the simulated galaxy population to reproduce the $z\sim0$ mass-metallicity relation. This is by no means a unique solution, but rather an illustrative example of how 
multiple observables are likely needed to constrain the underlying physical picture of galaxy formation and inform prescriptions for subgrid physics. \rrr{In broad and more speculative strokes, these results may indicate that the subgalactic-scale ISM/CGM dynamics and the relative roles of ejective and preventative star formation feedback play important roles in shaping the evolution of the low-mass galaxy population \citep[e.g.,][]{pandya2021, carr2022,fielding2022, pandya2023, smith2024}.}

The divergence in SFS redshift evolution between the cosmological simulations and observations demonstrates that low-redshift samples of dwarfs have the potential to constrain models of galaxy evolution beyond the oft-used $z=0$ scaling relations.  In particular, although all three theoretical works successfully reproduce the $z\sim0$ SFS within the work-to-work scatter of the observational literature, we see that the two cosmological simulations predict a significantly shallower \textit{evolution} in the  SFS as compared to both the work at hand and the broader observational literature.

\section{Conclusions}
In this work, we use the background galaxy spectra collected by the Satellites Around Galactic Analogs (SAGA) Survey to construct the \sagabg{} background sample and measure the low-redshift evolution of the low-mass
star-forming sequence (SFS). 
A quantitative measure of the cosmic evolution of low-mass galaxies is crucial both for understanding how dwarf galaxy populations are assembled over cosmic time and for using low-redshift galaxy samples to contextualize galaxies in the Nearby Universe.

We measured stellar masses and star formation rates for \nfinal{} galaxies with $z<0.21$ and 
\logmstar[$<10$] to assemble a picture of the evolution of the star-forming sequence over the last \tlbmax{}. By combining 
a simple model star formation rate and stellar mass accumulation with a data-driven model of our sample selection function,
we can disentangle the physical and observational contributions to this apparent evolution. We find a significant change in the normalization of the SFS from $z=0$ to $z=0.21$ down to \logmstar[$\sim 8$].
Notably, significant evolution in the SFS is detectable down to $z=0.05$, where we rule out zero evolution with 95\% confidence. The power-law index and intrinsic scatter of the SFS do not meaningfully evolve from $z=0$ to $z=0.21$. 
\rrr{The evolution of the SFS over this redshift range is similar in magnitude to the level of systematic uncertainty incurred from comparing different tracers of the star formation rate and stellar mass esimation methods: we thus suggest that precise measurements of the SFS should take into account both redshift-dependent and tracer-dependent observational systematics when measuring SFS redshift evolution or comparing to the Nearby Universe SFS.}

Our findings indicate that the physical evolution of the SFS inferred from the \sagabg{} sample aligns well with extrapolations of SFS fits from the observational literature, but shows significantly steeper evolution at low stellar mass than several contemporary cosmological hydrodynamical simulations. This implies that low-mass galaxies were, on average, more vigorously star-forming at low redshift than cosmological simulations currently predict, highlighting the need for improved models to accurately represent low-mass galaxy assembly.

A low redshift evolution of the SFS has two main implications: 
\begin{enumerate}
\item At the level of precision of \sagabg{}, including $z>0.05$ galaxies in ``$z\sim0$'' will confound redshift evolution with galactic physics (e.g., environmental effects of the Local Group) when comparing the SFS of Nearby galaxies and ``$z\sim 0$'' galaxies.
\item The low-redshift evolution of low-mass galaxies can test the evolution of the low-mass galaxy population. Although the empirical model \umdr{} matches our observed SFS evolution, both cosmological hydrodynamical simulations compared here (\tng{} and SIMBA) show significantly shallower evolution in the average SFR at fixed stellar mass.
\end{enumerate}
 
The redshift at which we rule out zero evolution in the SFS is significantly lower than common choices to define the local Universe for samples of similar or larger volume \citep[which can go up to $z\sim 0.3$, see e.g.,][]{salim2007, peng2010, renzini2015, battisti2017, kadofong2018, salim2018}.  As our measurements of both the Nearby Universe and the low-redshift Universe become increasingly more precise, it becomes increasingly more important to disentangle the physical origin of differences between the statistical properties of Nearby and low-redshift galaxies. Furthermore, precise measurements of 
the low-redshift Universe enables meaningful constraints on low-mass galaxy assembly over cosmic time, even when the dynamic range over lookback time is limited.

Surveys of dwarf galaxies in the low-redshift Universe (and beyond) are set to grow in 
mass completeness and volume at a rapid pace over the next decade \citep{darraghford2022, luo2023, sagaiii}. In order to leverage the information content of such surveys, it 
is imperative that we consider the physical implication of the increased precision that 
such surveys afford. This work represents a first but not all-encompassing effort to address the repercussions and utility of this progress. 

\acknowledgements{}
EKF gratefully acknowledges support from the YCAA Prize Postdoctoral Fellowship.
MG and YA were supported in part by a grant to MG~from the Howard Hughes Medical Institute (HHMI) through the HHMI Professors Program. Additional support was provided by the Kavli Institute for Particle Astrophysics and Cosmology at Stanford.

This research used data from the SAGA Survey (Satellites Around Galactic Analogs; sagasurvey.org). The SAGA Survey is a galaxy redshift survey with spectroscopic data obtained by the SAGA Survey team with the Anglo-Australian Telescope, MMT Observatory, Palomar Observatory, W. M. Keck Observatory, and the South African Astronomical Observatory (SAAO). The SAGA Survey also made use of many public data sets, including imaging data from the Sloan Digital Sky Survey (SDSS), the Dark Energy Survey (DES), the GALEX Survey, and the Dark Energy Spectroscopic Instrument (DESI) Legacy Imaging Surveys, which includes the Dark Energy Camera Legacy Survey (DECaLS), the Beijing-Arizona Sky Survey (BASS), and the Mayall z-band Legacy Survey (MzLS); 
redshift catalogs from SDSS, DESI, the Galaxy And Mass Assembly (GAMA) Survey, the Prism Multi-object Survey (PRIMUS), the VIMOS Public Extragalactic Redshift Survey (VIPERS), the WiggleZ Dark Energy Survey (WiggleZ), the 2dF Galaxy Redshift Survey (2dFGRS), the HectoMAP Redshift Survey, the HETDEX Source Catalog, the 6dF Galaxy Survey (6dFGS), the Hectospec Cluster Survey (HeCS), the Australian Dark Energy Survey (OzDES), the 2-degree Field Lensing Survey (2dFLenS), 
and the Las Campanas Redshift Survey (LCRS); HI data from the Arecibo Legacy Fast ALFA Survey (ALFALFA), the FAST all-sky HI Survey (FASHI), and HI Parkes All-Sky Survey (HIPASS); and compiled data from the NASA-Sloan Atlas (NSA), the Siena Galaxy Atlas (SGA), the HyperLeda database, and the Extragalactic Distance Database (EDD). The SAGA Survey was supported in part by NSF collaborative grants AST-1517148 and AST-1517422 and Heising–Simons Foundation grant 2019-1402. SAGA Survey’s full acknowledgments can be found at \url{https://sagasurvey.org/ack}.

\software{Astropy \citep{astropy:2013, astropy:2018}, matplotlib \citep{Hunter:2007}, SciPy \citep{scipy2020}, the IPython package \citep{PER-GRA:2007}, NumPy \citep{van2011numpy}, 
pandas \citep{pandas2022},
Astroquery \citep{astroquery}, extinction \citep{barbary2021}}

\bibliography{sagabg.bib,sagabg_paperii.bib,software.bib,gama.bib,SFMS_refs.bib}

\begin{thebibliography}{}
\expandafter\ifx\csname natexlab\endcsname\relax\def\natexlab#1{#1}\fi
\providecommand{\url}[1]{\href{#1}{#1}}
\providecommand{\dodoi}[1]{doi:~\href{http://doi.org/#1}{\nolinkurl{#1}}}
\providecommand{\doeprint}[1]{\href{http://ascl.net/#1}{\nolinkurl{http://ascl.net/#1}}}
\providecommand{\doarXiv}[1]{\href{https://arxiv.org/abs/#1}{\nolinkurl{https://arxiv.org/abs/#1}}}

\bibitem[{{Abolfathi} {et~al.}(2018){Abolfathi}, {Aguado}, {Aguilar}, {Allende
  Prieto}, {Almeida}, {Ananna}, {Anders}, {Anderson}, {Andrews}, {Anguiano},
  {Arag{\'o}n-Salamanca}, {Argudo-Fern{\'a}ndez}, {Armengaud}, {Ata},
  {Aubourg}, {Avila-Reese}, {Badenes}, {Bailey}, {Balland}, {Barger},
  {Barrera-Ballesteros}, {Bartosz}, {Bastien}, {Bates}, {Baumgarten},
  {Bautista}, {Beaton}, {Beers}, {Belfiore}, {Bender}, {Bernardi}, {Bershady},
  {Beutler}, {Bird}, {Bizyaev}, {Blanc}, {Blanton}, {Blomqvist}, {Bolton},
  {Boquien}, {Borissova}, {Bovy}, {Bradna Diaz}, {Brandt}, {Brinkmann},
  {Brownstein}, {Bundy}, {Burgasser}, {Burtin}, {Busca}, {Ca{\~n}as},
  {Cano-D{\'\i}az}, {Cappellari}, {Carrera}, {Casey}, {Cervantes Sodi}, {Chen},
  {Cherinka}, {Chiappini}, {Choi}, {Chojnowski}, {Chuang}, {Chung}, {Clerc},
  {Cohen}, {Comerford}, {Comparat}, {Correa do Nascimento}, {da Costa},
  {Cousinou}, {Covey}, {Crane}, {Cruz-Gonzalez}, {Cunha}, {da Silva Ilha},
  {Damke}, {Darling}, {Davidson}, {Dawson}, {de Icaza Lizaola}, {de la
  Macorra}, {de la Torre}, {De Lee}, {de Sainte Agathe}, {Deconto Machado},
  {Dell'Agli}, {Delubac}, {Diamond-Stanic}, {Donor}, {Downes}, {Drory}, {du Mas
  des Bourboux}, {Duckworth}, {Dwelly}, {Dyer}, {Ebelke}, {Davis Eigenbrot},
  {Eisenstein}, {Elsworth}, {Emsellem}, {Eracleous}, {Erfanianfar},
  {Escoffier}, {Fan}, {Fern{\'a}ndez Alvar}, {Fernandez-Trincado}, {Fernando
  Cirolini}, {Feuillet}, {Finoguenov}, {Fleming}, {Font-Ribera}, {Freischlad},
  {Frinchaboy}, {Fu}, {G{\'o}mez Maqueo Chew}, {Galbany}, {Garc{\'\i}a
  P{\'e}rez}, {Garcia-Dias}, {Garc{\'\i}a-Hern{\'a}ndez}, {Garma Oehmichen},
  {Gaulme}, {Gelfand}, {Gil-Mar{\'\i}n}, {Gillespie}, {Goddard}, {Gonz{\'a}lez
  Hern{\'a}ndez}, {Gonzalez-Perez}, {Grabowski}, {Green}, {Grier}, {Gueguen},
  {Guo}, {Guy}, {Hagen}, {Hall}, {Harding}, {Hasselquist}, {Hawley}, {Hayes},
  {Hearty}, {Hekker}, {Hernandez}, {Hernandez Toledo}, {Hogg},
  {Holley-Bockelmann}, {Holtzman}, {Hou}, {Hsieh}, {Hunt}, {Hutchinson},
  {Hwang}, {Jimenez Angel}, {Johnson}, {Jones}, {J{\"o}nsson}, {Jullo}, {Khan},
  {Kinemuchi}, {Kirkby}, {Kirkpatrick}, {Kitaura}, {Knapp}, {Kneib},
  {Kollmeier}, {Lacerna}, {Lane}, {Lang}, {Law}, {Le Goff}, {Lee}, {Li}, {Li},
  {Lian}, {Liang}, {Lima}, {Lin}, {Long}, {Lucatello}, {Lundgren}, {Mackereth},
  {MacLeod}, {Mahadevan}, {Maia}, {Majewski}, {Manchado}, {Maraston},
  {Mariappan}, {Marques-Chaves}, {Masseron}, {Masters}, {McDermid}, {McGreer},
  {Melendez}, {Meneses-Goytia}, {Merloni}, {Merrifield}, {Meszaros}, {Meza},
  {Minchev}, {Minniti}, {Mueller}, {Muller-Sanchez}, {Muna}, {Mu{\~n}oz},
  {Myers}, {Nair}, {Nandra}, {Ness}, {Newman}, {Nichol}, {Nidever},
  {Nitschelm}, {Noterdaeme}, {O'Connell}, {Oelkers}, {Oravetz}, {Oravetz},
  {Ort{\'\i}z}, {Osorio}, {Pace}, {Padilla}, {Palanque-Delabrouille},
  {Palicio}, {Pan}, {Pan}, {Parikh}, {P{\^a}ris}, {Park}, {Peirani},
  {Pellejero-Ibanez}, {Penny}, {Percival}, {Perez-Fournon}, {Petitjean},
  {Pieri}, {Pinsonneault}, {Pisani}, {Prada}, {Prakash}, {Queiroz}, {Raddick},
  {Raichoor}, {Barboza Rembold}, {Richstein}, {Riffel}, {Riffel}, {Rix},
  {Robin}, {Rodr{\'\i}guez Torres}, {Rom{\'a}n-Z{\'u}{\~n}iga}, {Ross},
  {Rossi}, {Ruan}, {Ruggeri}, {Ruiz}, {Salvato}, {S{\'a}nchez}, {S{\'a}nchez},
  {Sanchez Almeida}, {S{\'a}nchez-Gallego}, {Santana Rojas}, {Santiago},
  {Schiavon}, {Schimoia}, {Schlafly}, {Schlegel}, {Schneider}, {Schuster},
  {Schwope}, {Seo}, {Serenelli}, {Shen}, {Shen}, {Shetrone}, {Shull}, {Silva
  Aguirre}, {Simon}, {Skrutskie}, {Slosar}, {Smethurst}, {Smith}, {Sobeck},
  {Somers}, {Souter}, {Souto}, {Spindler}, {Stark}, {Stassun}, {Steinmetz},
  {Stello}, {Storchi-Bergmann}, {Streblyanska}, {Stringfellow}, {Su{\'a}rez},
  {Sun}, {Szigeti}, {Taghizadeh-Popp}, {Talbot}, {Tang}, {Tao}, {Tayar},
  {Tembe}, {Teske}, {Thakar}, {Thomas}, {Tissera}, {Tojeiro}, {Tremonti},
  {Troup}, {Urry}, {Valenzuela}, {van den Bosch}, {Vargas-Gonz{\'a}lez},
  {Vargas-Maga{\~n}a}, {Vazquez}, {Villanova}, {Vogt}, {Wake}, {Wang},
  {Weaver}, {Weijmans}, {Weinberg}, {Westfall}, {Whelan}, {Wilcots}, {Wild},
  {Williams}, {Wilson}, {Wood-Vasey}, {Wylezalek}, {Xiao}, {Yan}, {Yang},
  {Ybarra}, {Y{\`e}che}, {Zakamska}, {Zamora}, {Zarrouk}, {Zasowski}, {Zhang},
  {Zhao}, {Zhao}, {Zheng}, {Zheng}, {Zhou}, {Zhu}, {Zinn}, \&
  {Zou}}]{abolfathi2018}
{Abolfathi}, B., {Aguado}, D.~S., {Aguilar}, G., {et~al.} 2018, \apjs, 235, 42,
  \dodoi{10.3847/1538-4365/aa9e8a}

\bibitem[{{Andrews} \& {Martini}(2013)}]{andrews2013}
{Andrews}, B.~H., \& {Martini}, P. 2013, \apj, 765, 140,
  \dodoi{10.1088/0004-637X/765/2/140}

\bibitem[{{Astropy Collaboration} {et~al.}(2013){Astropy Collaboration},
  {Robitaille}, {Tollerud}, {Greenfield}, {Droettboom}, {Bray}, {Aldcroft},
  {Davis}, {Ginsburg}, {Price-Whelan}, {Kerzendorf}, {Conley}, {Crighton},
  {Barbary}, {Muna}, {Ferguson}, {Grollier}, {Parikh}, {Nair}, {Unther},
  {Deil}, {Woillez}, {Conseil}, {Kramer}, {Turner}, {Singer}, {Fox}, {Weaver},
  {Zabalza}, {Edwards}, {Azalee Bostroem}, {Burke}, {Casey}, {Crawford},
  {Dencheva}, {Ely}, {Jenness}, {Labrie}, {Lim}, {Pierfederici}, {Pontzen},
  {Ptak}, {Refsdal}, {Servillat}, \& {Streicher}}]{astropy:2013}
{Astropy Collaboration}, {Robitaille}, T.~P., {Tollerud}, E.~J., {et~al.} 2013,
  \aap, 558, A33, \dodoi{10.1051/0004-6361/201322068}

\bibitem[{{Baldry} {et~al.}(2010){Baldry}, {Robotham}, {Hill}, {Driver},
  {Liske}, {Norberg}, {Bamford}, {Hopkins}, {Loveday}, {Peacock}, {Cameron},
  {Croom}, {Cross}, {Doyle}, {Dye}, {Frenk}, {Jones}, {van Kampen}, {Kelvin},
  {Nichol}, {Parkinson}, {Popescu}, {Prescott}, {Sharp}, {Sutherland},
  {Thomas}, \& {Tuffs}}]{baldry2010}
{Baldry}, I.~K., {Robotham}, A.~S.~G., {Hill}, D.~T., {et~al.} 2010, \mnras,
  404, 86, \dodoi{10.1111/j.1365-2966.2010.16282.x}

\bibitem[{{Baldry} {et~al.}(2018){Baldry}, {Liske}, {Brown}, {Robotham},
  {Driver}, {Dunne}, {Alpaslan}, {Brough}, {Cluver}, {Eardley}, {Farrow},
  {Heymans}, {Hildebrandt}, {Hopkins}, {Kelvin}, {Loveday}, {Moffett},
  {Norberg}, {Owers}, {Taylor}, {Wright}, {Bamford}, {Bland-Hawthorn},
  {Bourne}, {Bremer}, {Colless}, {Conselice}, {Croom}, {Davies}, {Foster},
  {Grootes}, {Holwerda}, {Jones}, {Kafle}, {Kuijken}, {Lara-Lopez},
  {L{\'o}pez-S{\'a}nchez}, {Meyer}, {Phillipps}, {Sutherland}, {van Kampen}, \&
  {Wilkins}}]{baldry2018}
{Baldry}, I.~K., {Liske}, J., {Brown}, M.~J.~I., {et~al.} 2018, \mnras, 474,
  3875, \dodoi{10.1093/mnras/stx3042}

\bibitem[{{Barbary}(2021)}]{barbary2021}
{Barbary}, K. 2021, {extinction: Dust extinction laws}, Astrophysics Source
  Code Library, record ascl:2102.026.
\newblock \doeprint{2102.026}

\bibitem[{{Battisti} {et~al.}(2017){Battisti}, {Calzetti}, \&
  {Chary}}]{battisti2017}
{Battisti}, A.~J., {Calzetti}, D., \& {Chary}, R.~R. 2017, \apj, 840, 109,
  \dodoi{10.3847/1538-4357/aa6fb2}

\bibitem[{{Behroozi} {et~al.}(2019){Behroozi}, {Wechsler}, {Hearin}, \&
  {Conroy}}]{behroozi2019}
{Behroozi}, P., {Wechsler}, R.~H., {Hearin}, A.~P., \& {Conroy}, C. 2019,
  \mnras, 488, 3143, \dodoi{10.1093/mnras/stz1182}

\bibitem[{{Besla} {et~al.}(2018){Besla}, {Patton}, {Stierwalt},
  {Rodriguez-Gomez}, {Patel}, {Kallivayalil}, {Johnson}, {Pearson}, {Privon},
  \& {Putman}}]{besla2018}
{Besla}, G., {Patton}, D.~R., {Stierwalt}, S., {et~al.} 2018, \mnras, 480,
  3376, \dodoi{10.1093/mnras/sty2041}

\bibitem[{{Bolton} {et~al.}(2007){Bolton}, {Burles}, {Treu}, {Koopmans}, \&
  {Moustakas}}]{bolton2007}
{Bolton}, A.~S., {Burles}, S., {Treu}, T., {Koopmans}, L. V.~E., \&
  {Moustakas}, L.~A. 2007, \apjl, 665, L105, \dodoi{10.1086/521357}

\bibitem[{{Brinchmann} {et~al.}(2004){Brinchmann}, {Charlot}, {White},
  {Tremonti}, {Kauffmann}, {Heckman}, \& {Brinkmann}}]{brinchmann2004}
{Brinchmann}, J., {Charlot}, S., {White}, S.~D.~M., {et~al.} 2004, \mnras, 351,
  1151, \dodoi{10.1111/j.1365-2966.2004.07881.x}

\bibitem[{{Brook} {et~al.}(2014){Brook}, {Di Cintio}, {Knebe}, {Gottl{\"o}ber},
  {Hoffman}, {Yepes}, \& {Garrison-Kimmel}}]{brook2014}
{Brook}, C.~B., {Di Cintio}, A., {Knebe}, A., {et~al.} 2014, \apjl, 784, L14,
  \dodoi{10.1088/2041-8205/784/1/L14}

\bibitem[{{Brooks} {et~al.}(2007){Brooks}, {Governato}, {Booth}, {Willman},
  {Gardner}, {Wadsley}, {Stinson}, \& {Quinn}}]{brooks2007}
{Brooks}, A.~M., {Governato}, F., {Booth}, C.~M., {et~al.} 2007, \apjl, 655,
  L17, \dodoi{10.1086/511765}

\bibitem[{{Calzetti}(2013)}]{calzetti2013}
{Calzetti}, D. 2013, in Secular Evolution of Galaxies, ed.
  J.~{Falc{\'o}n-Barroso} \& J.~H. {Knapen}, 419,
  \dodoi{10.48550/arXiv.1208.2997}

\bibitem[{{Cano-D{\'\i}az} {et~al.}(2016){Cano-D{\'\i}az}, {S{\'a}nchez},
  {Zibetti}, {Ascasibar}, {Bland-Hawthorn}, {Ziegler}, {Gonz{\'a}lez Delgado},
  {Walcher}, {Garc{\'\i}a-Benito}, {Mast}, {Mendoza-P{\'e}rez},
  {Falc{\'o}n-Barroso}, {Galbany}, {Husemann}, {Kehrig}, {Marino},
  {S{\'a}nchez-Bl{\'a}zquez}, {L{\'o}pez-Cob{\'a}}, {L{\'o}pez-S{\'a}nchez}, \&
  {Vilchez}}]{canodiaz2016}
{Cano-D{\'\i}az}, M., {S{\'a}nchez}, S.~F., {Zibetti}, S., {et~al.} 2016,
  \apjl, 821, L26, \dodoi{10.3847/2041-8205/821/2/L26}

\bibitem[{{Carlsten} {et~al.}(2021){Carlsten}, {Greene}, {Peter}, {Beaton}, \&
  {Greco}}]{carlsten2021}
{Carlsten}, S.~G., {Greene}, J.~E., {Peter}, A. H.~G., {Beaton}, R.~L., \&
  {Greco}, J.~P. 2021, \apj, 908, 109, \dodoi{10.3847/1538-4357/abd039}

\bibitem[{Carr {et~al.}(2022)Carr, Bryan, Fielding, Pandya, \&
  Somerville}]{carr2022}
Carr, C., Bryan, G.~L., Fielding, D.~B., Pandya, V., \& Somerville, R.~S. 2022,
  Regulation of {Star} {Formation} by a {Hot} {Circumgalactic} {Medium}, Tech.
  rep., \dodoi{10.48550/arXiv.2211.05115}

\bibitem[{{Conroy} \& {Wechsler}(2009)}]{conroywechsler2009}
{Conroy}, C., \& {Wechsler}, R.~H. 2009, \apj, 696, 620,
  \dodoi{10.1088/0004-637X/696/1/620}

\bibitem[{{Cook} {et~al.}(2014){Cook}, {Dale}, {Johnson}, {Van Zee}, {Lee},
  {Kennicutt}, {Calzetti}, {Staudaher}, \& {Engelbracht}}]{cook2014}
{Cook}, D.~O., {Dale}, D.~A., {Johnson}, B.~D., {et~al.} 2014, \mnras, 445,
  899, \dodoi{10.1093/mnras/stu1787}

\bibitem[{{Daddi} {et~al.}(2022){Daddi}, {Delvecchio}, {Dimauro}, {Magnelli},
  {Gomez-Guijarro}, {Coogan}, {Elbaz}, {Kalita}, {Le Bail}, {Rich}, \&
  {Tan}}]{daddi2022}
{Daddi}, E., {Delvecchio}, I., {Dimauro}, P., {et~al.} 2022, \aap, 661, L7,
  \dodoi{10.1051/0004-6361/202243574}

\bibitem[{{Dale} {et~al.}(2023){Dale}, {Boquien}, {Turner}, {Calzetti},
  {Kennicutt}, \& {Lee}}]{dale2023}
{Dale}, D.~A., {Boquien}, M., {Turner}, J.~A., {et~al.} 2023, \aj, 165, 260,
  \dodoi{10.3847/1538-3881/accffe}

\bibitem[{{Dale} {et~al.}(2009){Dale}, {Cohen}, {Johnson}, {Schuster},
  {Calzetti}, {Engelbracht}, {Gil de Paz}, {Kennicutt}, {Lee}, {Begum},
  {Block}, {Dalcanton}, {Funes}, {Gordon}, {Johnson}, {Marble}, {Sakai},
  {Skillman}, {van Zee}, {Walter}, {Weisz}, {Williams}, {Wu}, \&
  {Wu}}]{dale2009}
{Dale}, D.~A., {Cohen}, S.~A., {Johnson}, L.~C., {et~al.} 2009, \apj, 703, 517,
  \dodoi{10.1088/0004-637X/703/1/517}

\bibitem[{{Darragh-Ford} {et~al.}(2022){Darragh-Ford}, {Wu}, {Mao}, {Wechsler},
  {Geha}, {Forero-Romero}, {Hahn}, {Kallivayalil}, {Moustakas}, {Nadler},
  {Nowotka}, {Peek}, {Tollerud}, {Weiner}, {Aguilar}, {Ahlen}, {Brooks},
  {Cooper}, {de la Macorra}, {Dey}, {Fanning}, {Font-Ribera}, {Gontcho},
  {Honscheid}, {Kisner}, {Kremin}, {Landriau}, {Levi}, {Martini}, {Meisner},
  {Miquel}, {Myers}, {Nie}, {Palanque-Delabrouille}, {Percival}, {Prada},
  {Schlegel}, {Schubnell}, {Tarl{\'e}}, {Vargas-Maga{\~n}a}, {Zhou}, \&
  {Zou}}]{darraghford2022}
{Darragh-Ford}, E., {Wu}, J.~F., {Mao}, Y.-Y., {et~al.} 2022, arXiv e-prints,
  arXiv:2212.07433, \dodoi{10.48550/arXiv.2212.07433}

\bibitem[{{Dav{\'e}} {et~al.}(2019){Dav{\'e}}, {Angl{\'e}s-Alc{\'a}zar},
  {Narayanan}, {Li}, {Rafieferantsoa}, \& {Appleby}}]{dave2019}
{Dav{\'e}}, R., {Angl{\'e}s-Alc{\'a}zar}, D., {Narayanan}, D., {et~al.} 2019,
  \mnras, 486, 2827, \dodoi{10.1093/mnras/stz937}

\bibitem[{{Davies} {et~al.}(2016){Davies}, {Driver}, {Robotham}, {Grootes},
  {Popescu}, {Tuffs}, {Hopkins}, {Alpaslan}, {Andrews}, {Bland-Hawthorn},
  {Bremer}, {Brough}, {Brown}, {Cluver}, {Croom}, {da Cunha}, {Dunne},
  {Lara-L{\'o}pez}, {Liske}, {Loveday}, {Moffett}, {Owers}, {Phillipps},
  {Sansom}, {Taylor}, {Michalowski}, {Ibar}, {Smith}, \& {Bourne}}]{davies2016}
{Davies}, L.~J.~M., {Driver}, S.~P., {Robotham}, A.~S.~G., {et~al.} 2016,
  \mnras, 461, 458, \dodoi{10.1093/mnras/stw1342}

\bibitem[{{de los Reyes} \& {Kennicutt}(2019)}]{delosreyes2019}
{de los Reyes}, M. A.~C., \& {Kennicutt}, Robert~C., J. 2019, \apj, 872, 16,
  \dodoi{10.3847/1538-4357/aafa82}

\bibitem[{{Dekel} \& {Silk}(1986)}]{dekelsilk1986}
{Dekel}, A., \& {Silk}, J. 1986, \apj, 303, 39, \dodoi{10.1086/164050}

\bibitem[{{Dey} {et~al.}(2022){Dey}, {Andrews}, {Newman}, {Mao}, {Rau}, \&
  {Zhou}}]{dey2022}
{Dey}, B., {Andrews}, B.~H., {Newman}, J.~A., {et~al.} 2022, \mnras, 515, 5285,
  \dodoi{10.1093/mnras/stac2105}

\bibitem[{{El-Badry} {et~al.}(2016){El-Badry}, {Wetzel}, {Geha}, {Hopkins},
  {Kere{\v{s}}}, {Chan}, \& {Faucher-Gigu{\`e}re}}]{elbadry2016}
{El-Badry}, K., {Wetzel}, A., {Geha}, M., {et~al.} 2016, \apj, 820, 131,
  \dodoi{10.3847/0004-637X/820/2/131}

\bibitem[{{El-Badry} {et~al.}(2018){El-Badry}, {Bradford}, {Quataert}, {Geha},
  {Boylan-Kolchin}, {Weisz}, {Wetzel}, {Hopkins}, {Chan}, {Fitts},
  {Kere{\v{s}}}, \& {Faucher-Gigu{\`e}re}}]{elbadry2018}
{El-Badry}, K., {Bradford}, J., {Quataert}, E., {et~al.} 2018, \mnras, 477,
  1536, \dodoi{10.1093/mnras/sty730}

\bibitem[{{Ellison} {et~al.}(2020){Ellison}, {Thorp}, {Lin}, {Pan}, {Bluck},
  {Scudder}, {Teimoorinia}, {S{\'a}nchez}, \& {Sargent}}]{ellison2020}
{Ellison}, S.~L., {Thorp}, M.~D., {Lin}, L., {et~al.} 2020, \mnras, 493, L39,
  \dodoi{10.1093/mnrasl/slz179}

\bibitem[{{Ellison} {et~al.}(2024){Ellison}, {Pan}, {Bluck}, {Krumholz}, {Lin},
  {Hunt}, {Corbelli}, {Thorp}, {Barrera-Ballesteros}, {S{\'a}nchez}, {Scudder},
  \& {Quai}}]{ellison2024}
{Ellison}, S.~L., {Pan}, H.-A., {Bluck}, A. F.~L., {et~al.} 2024, \mnras, 527,
  10201, \dodoi{10.1093/mnras/stad3778}

\bibitem[{{Faber} \& {Jackson}(1976)}]{faber1976}
{Faber}, S.~M., \& {Jackson}, R.~E. 1976, \apj, 204, 668,
  \dodoi{10.1086/154215}

\bibitem[{{Ferrarese} {et~al.}(2001){Ferrarese}, {Pogge}, {Peterson},
  {Merritt}, {Wandel}, \& {Joseph}}]{ferrarese2001}
{Ferrarese}, L., {Pogge}, R.~W., {Peterson}, B.~M., {et~al.} 2001, \apjl, 555,
  L79, \dodoi{10.1086/322528}

\bibitem[{{Fielding} \& {Bryan}(2022)}]{fielding2022}
{Fielding}, D.~B., \& {Bryan}, G.~L. 2022, \apj, 924, 82,
  \dodoi{10.3847/1538-4357/ac2f41}

\bibitem[{{Foreman-Mackey} {et~al.}(2013){Foreman-Mackey}, {Hogg}, {Lang}, \&
  {Goodman}}]{emcee}
{Foreman-Mackey}, D., {Hogg}, D.~W., {Lang}, D., \& {Goodman}, J. 2013, \pasp,
  125, 306, \dodoi{10.1086/670067}

\bibitem[{{Geha} {et~al.}(2012){Geha}, {Blanton}, {Yan}, \&
  {Tinker}}]{geha2012}
{Geha}, M., {Blanton}, M.~R., {Yan}, R., \& {Tinker}, J.~L. 2012, \apj, 757,
  85, \dodoi{10.1088/0004-637X/757/1/85}

\bibitem[{{Geha} {et~al.}(2017){Geha}, {Wechsler}, {Mao}, {Tollerud}, {Weiner},
  {Bernstein}, {Hoyle}, {Marchi}, {Marshall}, {Mu{\~n}oz}, \& {Lu}}]{geha2017}
{Geha}, M., {Wechsler}, R.~H., {Mao}, Y.-Y., {et~al.} 2017, \apj, 847, 4,
  \dodoi{10.3847/1538-4357/aa8626}

\bibitem[{{Geha} {et~al.}(2024){Geha}, {Mao}, {Wechsler}, {Asali}, {Kado-Fong},
  {Kallivayalil}, {Nadler}, {Tollerud}, {Weiner}, {de los Reyes}, {Wang}, \&
  {Wu}}]{sagaiv}
{Geha}, M., {Mao}, Y.-Y., {Wechsler}, R.~H., {et~al.} 2024, arXiv e-prints,
  arXiv:2404.14499, \dodoi{10.48550/arXiv.2404.14499}

\bibitem[{{Ginsburg} {et~al.}(2019){Ginsburg}, {Sip{\H{o}}cz}, {Brasseur},
  {Cowperthwaite}, {Craig}, {Deil}, {Guillochon}, {Guzman}, {Liedtke}, {Lian
  Lim}, {Lockhart}, {Mommert}, {Morris}, {Norman}, {Parikh}, {Persson},
  {Robitaille}, {Segovia}, {Singer}, {Tollerud}, {de Val-Borro}, {Valtchanov},
  {Woillez}, {Astroquery Collaboration}, \& {a subset of astropy
  Collaboration}}]{astroquery}
{Ginsburg}, A., {Sip{\H{o}}cz}, B.~M., {Brasseur}, C.~E., {et~al.} 2019, \aj,
  157, 98, \dodoi{10.3847/1538-3881/aafc33}

\bibitem[{{Gonz{\'a}lez Delgado} {et~al.}(1999){Gonz{\'a}lez Delgado},
  {Leitherer}, \& {Heckman}}]{gonzalezdelgado1999}
{Gonz{\'a}lez Delgado}, R.~M., {Leitherer}, C., \& {Heckman}, T.~M. 1999,
  \apjs, 125, 489, \dodoi{10.1086/313285}

\bibitem[{{Hahn} {et~al.}(2019){Hahn}, {Starkenburg}, {Choi}, {Dav{\'e}},
  {Dickey}, {Geha}, {Genel}, {Hayward}, {Maller}, {Mandyam}, {Pandya},
  {Popping}, {Rafieferantsoa}, {Somerville}, \& {Tinker}}]{hahn2019}
{Hahn}, C., {Starkenburg}, T.~K., {Choi}, E., {et~al.} 2019, \apj, 872, 160,
  \dodoi{10.3847/1538-4357/aafedd}

\bibitem[{{Hopkins} {et~al.}(2018){Hopkins}, {Wetzel}, {Kere{\v{s}}},
  {Faucher-Gigu{\`e}re}, {Quataert}, {Boylan-Kolchin}, {Murray}, {Hayward},
  {Garrison-Kimmel}, {Hummels}, {Feldmann}, {Torrey}, {Ma},
  {Angl{\'e}s-Alc{\'a}zar}, {Su}, {Orr}, {Schmitz}, {Escala}, {Sanderson},
  {Grudi{\'c}}, {Hafen}, {Kim}, {Fitts}, {Bullock}, {Wheeler}, {Chan},
  {Elbert}, \& {Narayanan}}]{hopkins2018}
{Hopkins}, P.~F., {Wetzel}, A., {Kere{\v{s}}}, D., {et~al.} 2018, \mnras, 480,
  800, \dodoi{10.1093/mnras/sty1690}

\bibitem[{Hunter(2007)}]{Hunter:2007}
Hunter, J.~D. 2007, Computing in Science \& Engineering, 9, 90,
  \dodoi{10.1109/MCSE.2007.55}

\bibitem[{{Jiang} {et~al.}(2021){Jiang}, {Dekel}, {Freundlich}, {van den
  Bosch}, {Green}, {Hopkins}, {Benson}, \& {Du}}]{jiang2021}
{Jiang}, F., {Dekel}, A., {Freundlich}, J., {et~al.} 2021, \mnras, 502, 621,
  \dodoi{10.1093/mnras/staa4034}

\bibitem[{{Kado-Fong} {et~al.}(2020{\natexlab{a}}){Kado-Fong}, {Greene},
  {Greco}, {Beaton}, {Goulding}, {Johnson}, \& {Komiyama}}]{kadofong2020a}
{Kado-Fong}, E., {Greene}, J.~E., {Greco}, J.~P., {et~al.} 2020{\natexlab{a}},
  \aj, 159, 103, \dodoi{10.3847/1538-3881/ab6ef3}

\bibitem[{{Kado-Fong} {et~al.}(2020{\natexlab{b}}){Kado-Fong}, {Greene},
  {Huang}, {Beaton}, {Goulding}, \& {Komiyama}}]{kadofong2020b}
{Kado-Fong}, E., {Greene}, J.~E., {Huang}, S., {et~al.} 2020{\natexlab{b}},
  \apj, 900, 163, \dodoi{10.3847/1538-4357/abacc2}

\bibitem[{{Kado-Fong} {et~al.}(2022{\natexlab{a}}){Kado-Fong}, {Kim}, {Greene},
  \& {Lancaster}}]{kadofong2022c}
{Kado-Fong}, E., {Kim}, C.-G., {Greene}, J.~E., \& {Lancaster}, L.
  2022{\natexlab{a}}, \apj, 939, 101, \dodoi{10.3847/1538-4357/ac9673}

\bibitem[{{Kado-Fong} {et~al.}(2018){Kado-Fong}, {Greene}, {Hendel},
  {Price-Whelan}, {Greco}, {Goulding}, {Huang}, {Johnston}, {Komiyama}, {Lee},
  {Lust}, {Strauss}, \& {Tanaka}}]{kadofong2018}
{Kado-Fong}, E., {Greene}, J.~E., {Hendel}, D., {et~al.} 2018, \apj, 866, 103,
  \dodoi{10.3847/1538-4357/aae0f0}

\bibitem[{{Kado-Fong} {et~al.}(2022{\natexlab{b}}){Kado-Fong}, {Sanderson},
  {Greene}, {Cunningham}, {Wheeler}, {Chan}, {El-Badry}, {Hopkins}, {Wetzel},
  {Boylan-Kolchin}, {Faucher-Gigu{\`e}re}, {Huang}, {Quataert}, \&
  {Starkenburg}}]{kadofong2022a}
{Kado-Fong}, E., {Sanderson}, R.~E., {Greene}, J.~E., {et~al.}
  2022{\natexlab{b}}, \apj, 931, 152, \dodoi{10.3847/1538-4357/ac6c88}

\bibitem[{{Kado-Fong} {et~al.}(2024){Kado-Fong}, {Geha}, {Mao}, {de los Reyes},
  {Wechsler}, {Asali}, {Kallivayalil}, {Nadler}, {Tollerud}, \&
  {Weiner}}]{kadofong2024}
{Kado-Fong}, E., {Geha}, M., {Mao}, Y.-Y., {et~al.} 2024, \apj, 966, 129,
  \dodoi{10.3847/1538-4357/ad3042}

\bibitem[{{Karim} {et~al.}(2011){Karim}, {Schinnerer},
  {Mart{\'\i}nez-Sansigre}, {Sargent}, {van der Wel}, {Rix}, {Ilbert},
  {Smol{\v{c}}i{\'c}}, {Carilli}, {Pannella}, {Koekemoer}, {Bell}, \&
  {Salvato}}]{karim2011}
{Karim}, A., {Schinnerer}, E., {Mart{\'\i}nez-Sansigre}, A., {et~al.} 2011,
  \apj, 730, 61, \dodoi{10.1088/0004-637X/730/2/61}

\bibitem[{{Kennicutt}(1989)}]{kennicutt1989}
{Kennicutt}, Robert~C., J. 1989, \apj, 344, 685, \dodoi{10.1086/167834}

\bibitem[{{Kennicutt}(1998)}]{kennicutt1998}
---. 1998, \apj, 498, 541, \dodoi{10.1086/305588}

\bibitem[{{Kormendy}(1985)}]{kormendy1985}
{Kormendy}, J. 1985, \apj, 295, 73, \dodoi{10.1086/163350}

\bibitem[{{Kroupa}(2001)}]{kroupa2001}
{Kroupa}, P. 2001, \mnras, 322, 231, \dodoi{10.1046/j.1365-8711.2001.04022.x}

\bibitem[{{Lee} {et~al.}(2009){Lee}, {Gil de Paz}, {Tremonti}, {Kennicutt},
  {Salim}, {Bothwell}, {Calzetti}, {Dalcanton}, {Dale}, {Engelbracht}, {Funes},
  {Johnson}, {Sakai}, {Skillman}, {van Zee}, {Walter}, \& {Weisz}}]{lee2009}
{Lee}, J.~C., {Gil de Paz}, A., {Tremonti}, C., {et~al.} 2009, \apj, 706, 599,
  \dodoi{10.1088/0004-637X/706/1/599}

\bibitem[{{Lee} {et~al.}(2015){Lee}, {Sanders}, {Casey}, {Toft}, {Scoville},
  {Hung}, {Le Floc'h}, {Ilbert}, {Zahid}, {Aussel}, {Capak}, {Kartaltepe},
  {Kewley}, {Li}, {Schawinski}, {Sheth}, \& {Xiao}}]{lee2015}
{Lee}, N., {Sanders}, D.~B., {Casey}, C.~M., {et~al.} 2015, \apj, 801, 80,
  \dodoi{10.1088/0004-637X/801/2/80}

\bibitem[{{Leja} {et~al.}(2022){Leja}, {Speagle}, {Ting}, {Johnson}, {Conroy},
  {Whitaker}, {Nelson}, {van Dokkum}, \& {Franx}}]{leja2022}
{Leja}, J., {Speagle}, J.~S., {Ting}, Y.-S., {et~al.} 2022, \apj, 936, 165,
  \dodoi{10.3847/1538-4357/ac887d}

\bibitem[{{Leroy} {et~al.}(2019){Leroy}, {Sandstrom}, {Lang}, {Lewis}, {Salim},
  {Behrens}, {Chastenet}, {Chiang}, {Gallagher}, {Kessler}, \&
  {Utomo}}]{leroy2019}
{Leroy}, A.~K., {Sandstrom}, K.~M., {Lang}, D., {et~al.} 2019, \apjs, 244, 24,
  \dodoi{10.3847/1538-4365/ab3925}

\bibitem[{{Leslie} {et~al.}(2020){Leslie}, {Schinnerer}, {Liu}, {Magnelli},
  {Algera}, {Karim}, {Davidzon}, {Gozaliasl}, {Jim{\'e}nez-Andrade}, {Lang},
  {Sargent}, {Novak}, {Groves}, {Smol{\v{c}}i{\'c}}, {Zamorani}, {Vaccari},
  {Battisti}, {Vardoulaki}, {Peng}, \& {Kartaltepe}}]{leslie2020}
{Leslie}, S.~K., {Schinnerer}, E., {Liu}, D., {et~al.} 2020, \apj, 899, 58,
  \dodoi{10.3847/1538-4357/aba044}

\bibitem[{{Lin} {et~al.}(2019){Lin}, {Pan}, {Ellison}, {Belfiore}, {Shi},
  {S{\'a}nchez}, {Hsieh}, {Rowlands}, {Ramya}, {Thorp}, {Li}, \&
  {Maiolino}}]{lin2019}
{Lin}, L., {Pan}, H.-A., {Ellison}, S.~L., {et~al.} 2019, \apjl, 884, L33,
  \dodoi{10.3847/2041-8213/ab4815}

\bibitem[{{Luo} {et~al.}(2023){Luo}, {Leauthaud}, {Greene}, {Huang},
  {Kado-Fong}, {Danieli}, {Li}, {Li}, {Blanco}, {Wasleske}, {Wick}, {Mintz},
  {Guan}, {Peter}, {Baldassare}, {Brooks}, {Banerjee}, {Bhattacharyya}, {Cai},
  {Chen}, {Gunn}, {Johnson}, {Kelvin}, {Li}, {Lin}, {Lupton}, {Mace}, {Medina},
  {Read}, {Cordova Rosado}, \& {Seifert}}]{luo2023}
{Luo}, Y., {Leauthaud}, A., {Greene}, J., {et~al.} 2023, arXiv e-prints,
  arXiv:2305.19310, \dodoi{10.48550/arXiv.2305.19310}

\bibitem[{{Mao} {et~al.}(2021){Mao}, {Geha}, {Wechsler}, {Weiner}, {Tollerud},
  {Nadler}, \& {Kallivayalil}}]{mao2021}
{Mao}, Y.-Y., {Geha}, M., {Wechsler}, R.~H., {et~al.} 2021, \apj, 907, 85,
  \dodoi{10.3847/1538-4357/abce58}

\bibitem[{{Mao} {et~al.}(2024){Mao}, {Geha}, {Wechsler}, {Asali}, {Wang},
  {Kado-Fong}, {Kallivayalil}, {Nadler}, {Tollerud}, {Weiner}, {de los Reyes},
  \& {Wu}}]{sagaiii}
---. 2024, arXiv e-prints, arXiv:2404.14498, \dodoi{10.48550/arXiv.2404.14498}

\bibitem[{{McQuinn} {et~al.}(2015){McQuinn}, {Cannon}, {Dolphin}, {Skillman},
  {Haynes}, {Simones}, {Salzer}, {Adams}, {Elson}, {Giovanelli}, \&
  {Ott}}]{mcquinn2015}
{McQuinn}, K. B.~W., {Cannon}, J.~M., {Dolphin}, A.~E., {et~al.} 2015, \apj,
  802, 66, \dodoi{10.1088/0004-637X/802/1/66}

\bibitem[{{Morselli} {et~al.}(2020){Morselli}, {Rodighiero}, {Enia},
  {Corbelli}, {Casasola}, {Rodr{\'\i}guez-Mu{\~n}oz}, {Renzini}, {Tacchella},
  {Baronchelli}, {Bianchi}, {Cassata}, {Franceschini}, {Mancini}, {Negrello},
  {Popesso}, \& {Romano}}]{morselli2020}
{Morselli}, L., {Rodighiero}, G., {Enia}, A., {et~al.} 2020, \mnras, 496, 4606,
  \dodoi{10.1093/mnras/staa1811}

\bibitem[{{Munshi} {et~al.}(2021){Munshi}, {Brooks}, {Applebaum},
  {Christensen}, {Quinn}, \& {Sligh}}]{munshi2021}
{Munshi}, F., {Brooks}, A.~M., {Applebaum}, E., {et~al.} 2021, \apj, 923, 35,
  \dodoi{10.3847/1538-4357/ac0db6}

\bibitem[{{Nelson} {et~al.}(2019){Nelson}, {Springel}, {Pillepich},
  {Rodriguez-Gomez}, {Torrey}, {Genel}, {Vogelsberger}, {Pakmor}, {Marinacci},
  {Weinberger}, {Kelley}, {Lovell}, {Diemer}, \& {Hernquist}}]{nelson2019}
{Nelson}, D., {Springel}, V., {Pillepich}, A., {et~al.} 2019, Computational
  Astrophysics and Cosmology, 6, 2, \dodoi{10.1186/s40668-019-0028-x}

\bibitem[{{Noeske} {et~al.}(2007){Noeske}, {Weiner}, {Faber}, {Papovich},
  {Koo}, {Somerville}, {Bundy}, {Conselice}, {Newman}, {Schiminovich}, {Le
  Floc'h}, {Coil}, {Rieke}, {Lotz}, {Primack}, {Barmby}, {Cooper}, {Davis},
  {Ellis}, {Fazio}, {Guhathakurta}, {Huang}, {Kassin}, {Martin}, {Phillips},
  {Rich}, {Small}, {Willmer}, \& {Wilson}}]{noeske2007}
{Noeske}, K.~G., {Weiner}, B.~J., {Faber}, S.~M., {et~al.} 2007, \apjl, 660,
  L43, \dodoi{10.1086/517926}

\bibitem[{{Ostriker} \& {Kim}(2022)}]{ostriker2022}
{Ostriker}, E.~C., \& {Kim}, C.-G. 2022, \apj, 936, 137,
  \dodoi{10.3847/1538-4357/ac7de2}

\bibitem[{{Pahre} {et~al.}(1998){Pahre}, {Djorgovski}, \& {de
  Carvalho}}]{pahre1998}
{Pahre}, M.~A., {Djorgovski}, S.~G., \& {de Carvalho}, R.~R. 1998, \aj, 116,
  1591, \dodoi{10.1086/300544}

\bibitem[{{Pandya} {et~al.}(2021){Pandya}, {Fielding},
  {Angl{\'e}s-Alc{\'a}zar}, {Somerville}, {Bryan}, {Hayward}, {Stern}, {Kim},
  {Quataert}, {Forbes}, {Faucher-Gigu{\`e}re}, {Feldmann}, {Hafen}, {Hopkins},
  {Kere{\v{s}}}, {Murray}, \& {Wetzel}}]{pandya2021}
{Pandya}, V., {Fielding}, D.~B., {Angl{\'e}s-Alc{\'a}zar}, D., {et~al.} 2021,
  \mnras, 508, 2979, \dodoi{10.1093/mnras/stab2714}

\bibitem[{{Pandya} {et~al.}(2023){Pandya}, {Fielding}, {Bryan}, {Carr},
  {Somerville}, {Stern}, {Faucher-Gigu{\`e}re}, {Hafen},
  {Angl{\'e}s-Alc{\'a}zar}, \& {Forbes}}]{pandya2023}
{Pandya}, V., {Fielding}, D.~B., {Bryan}, G.~L., {et~al.} 2023, \apj, 956, 118,
  \dodoi{10.3847/1538-4357/acf3ea}

\bibitem[{{Pannella} {et~al.}(2009){Pannella}, {Gabasch}, {Goranova}, {Drory},
  {Hopp}, {Noll}, {Saglia}, {Strazzullo}, \& {Bender}}]{pannella2009}
{Pannella}, M., {Gabasch}, A., {Goranova}, Y., {et~al.} 2009, \apj, 701, 787,
  \dodoi{10.1088/0004-637X/701/1/787}

\bibitem[{{Pannella} {et~al.}(2015){Pannella}, {Elbaz}, {Daddi}, {Dickinson},
  {Hwang}, {Schreiber}, {Strazzullo}, {Aussel}, {Bethermin}, {Buat},
  {Charmandaris}, {Cibinel}, {Juneau}, {Ivison}, {Le Borgne}, {Le Floc'h},
  {Leiton}, {Lin}, {Magdis}, {Morrison}, {Mullaney}, {Onodera}, {Renzini},
  {Salim}, {Sargent}, {Scott}, {Shu}, \& {Wang}}]{pannella2015}
{Pannella}, M., {Elbaz}, D., {Daddi}, E., {et~al.} 2015, \apj, 807, 141,
  \dodoi{10.1088/0004-637X/807/2/141}

\bibitem[{{Pasquet} {et~al.}(2019){Pasquet}, {Bertin}, {Treyer}, {Arnouts}, \&
  {Fouchez}}]{pasquet2019}
{Pasquet}, J., {Bertin}, E., {Treyer}, M., {Arnouts}, S., \& {Fouchez}, D.
  2019, \aap, 621, A26, \dodoi{10.1051/0004-6361/201833617}

\bibitem[{{Pearson} {et~al.}(2018){Pearson}, {Wang}, {Hurley}, {Ma{\l}ek},
  {Buat}, {Burgarella}, {Farrah}, {Oliver}, {Smith}, \& {van der
  Tak}}]{pearson2018}
{Pearson}, W.~J., {Wang}, L., {Hurley}, P.~D., {et~al.} 2018, \aap, 615, A146,
  \dodoi{10.1051/0004-6361/201832821}

\bibitem[{{Peng} {et~al.}(2010){Peng}, {Lilly}, {Kova{\v{c}}}, {Bolzonella},
  {Pozzetti}, {Renzini}, {Zamorani}, {Ilbert}, {Knobel}, {Iovino}, {Maier},
  {Cucciati}, {Tasca}, {Carollo}, {Silverman}, {Kampczyk}, {de Ravel},
  {Sanders}, {Scoville}, {Contini}, {Mainieri}, {Scodeggio}, {Kneib}, {Le
  F{\`e}vre}, {Bardelli}, {Bongiorno}, {Caputi}, {Coppa}, {de la Torre},
  {Franzetti}, {Garilli}, {Lamareille}, {Le Borgne}, {Le Brun}, {Mignoli},
  {Perez Montero}, {Pello}, {Ricciardelli}, {Tanaka}, {Tresse}, {Vergani},
  {Welikala}, {Zucca}, {Oesch}, {Abbas}, {Barnes}, {Bordoloi}, {Bottini},
  {Cappi}, {Cassata}, {Cimatti}, {Fumana}, {Hasinger}, {Koekemoer},
  {Leauthaud}, {Maccagni}, {Marinoni}, {McCracken}, {Memeo}, {Meneux}, {Nair},
  {Porciani}, {Presotto}, \& {Scaramella}}]{peng2010}
{Peng}, Y.-j., {Lilly}, S.~J., {Kova{\v{c}}}, K., {et~al.} 2010, \apj, 721,
  193, \dodoi{10.1088/0004-637X/721/1/193}

\bibitem[{P\'erez \& Granger(2007)}]{PER-GRA:2007}
P\'erez, F., \& Granger, B.~E. 2007, Computing in Science and Engineering, 9,
  21, \dodoi{10.1109/MCSE.2007.53}

\bibitem[{{Pessa} {et~al.}(2021){Pessa}, {Schinnerer}, {Belfiore}, {Emsellem},
  {Leroy}, {Schruba}, {Kruijssen}, {Pan}, {Blanc}, {Sanchez-Blazquez},
  {Bigiel}, {Chevance}, {Congiu}, {Dale}, {Faesi}, {Glover}, {Grasha},
  {Groves}, {Ho}, {Jim{\'e}nez-Donaire}, {Klessen}, {Kreckel}, {Koch}, {Liu},
  {Meidt}, {Pety}, {Querejeta}, {Rosolowsky}, {Saito}, {Santoro}, {Sun},
  {Usero}, {Watkins}, \& {Williams}}]{pessa2021}
{Pessa}, I., {Schinnerer}, E., {Belfiore}, F., {et~al.} 2021, \aap, 650, A134,
  \dodoi{10.1051/0004-6361/202140733}

\bibitem[{{Pessa} {et~al.}(2022){Pessa}, {Schinnerer}, {Leroy}, {Koch},
  {Rosolowsky}, {Williams}, {Pan}, {Schruba}, {Usero}, {Belfiore}, {Bigiel},
  {Blanc}, {Chevance}, {Dale}, {Emsellem}, {Gensior}, {Glover}, {Grasha},
  {Groves}, {Klessen}, {Kreckel}, {Kruijssen}, {Liu}, {Meidt}, {Pety},
  {Querejeta}, {Saito}, {Sanchez-Blazquez}, \& {Watkins}}]{pessa2022}
{Pessa}, I., {Schinnerer}, E., {Leroy}, A.~K., {et~al.} 2022, \aap, 663, A61,
  \dodoi{10.1051/0004-6361/202142832}

\bibitem[{{Popesso} {et~al.}(2019){Popesso}, {Morselli}, {Concas}, {Schreiber},
  {Rodighiero}, {Cresci}, {Belli}, {Ilbert}, {Erfanianfar}, {Mancini}, {Inami},
  {Dickinson}, {Pannella}, \& {Elbaz}}]{popesso2019}
{Popesso}, P., {Morselli}, L., {Concas}, A., {et~al.} 2019, \mnras, 490, 5285,
  \dodoi{10.1093/mnras/stz2635}

\bibitem[{{Popesso} {et~al.}(2023){Popesso}, {Concas}, {Cresci}, {Belli},
  {Rodighiero}, {Inami}, {Dickinson}, {Ilbert}, {Pannella}, \&
  {Elbaz}}]{popesso2023}
{Popesso}, P., {Concas}, A., {Cresci}, G., {et~al.} 2023, \mnras, 519, 1526,
  \dodoi{10.1093/mnras/stac3214}

\bibitem[{{Price-Whelan} {et~al.}(2018){Price-Whelan}, {Sip{\H{o}}cz},
  {G{\"u}nther}, {Lim}, {Crawford}, {Conseil}, {Shupe}, {Craig}, {Dencheva},
  {Ginsburg}, {VanderPlas}, {Bradley}, {P{\'e}rez-Su{\'a}rez}, {de Val-Borro},
  {Paper Contributors}, {Aldcroft}, {Cruz}, {Robitaille}, {Tollerud},
  {Coordination Committee}, {Ardelean}, {Babej}, {Bach}, {Bachetti}, {Bakanov},
  {Bamford}, {Barentsen}, {Barmby}, {Baumbach}, {Berry}, {Biscani}, {Boquien},
  {Bostroem}, {Bouma}, {Brammer}, {Bray}, {Breytenbach}, {Buddelmeijer},
  {Burke}, {Calderone}, {Cano Rodr{\'\i}guez}, {Cara}, {Cardoso}, {Cheedella},
  {Copin}, {Corrales}, {Crichton}, {D{\textquoteright}Avella}, {Deil},
  {Depagne}, {Dietrich}, {Donath}, {Droettboom}, {Earl}, {Erben}, {Fabbro},
  {Ferreira}, {Finethy}, {Fox}, {Garrison}, {Gibbons}, {Goldstein}, {Gommers},
  {Greco}, {Greenfield}, {Groener}, {Grollier}, {Hagen}, {Hirst}, {Homeier},
  {Horton}, {Hosseinzadeh}, {Hu}, {Hunkeler}, {Ivezi{\'c}}, {Jain}, {Jenness},
  {Kanarek}, {Kendrew}, {Kern}, {Kerzendorf}, {Khvalko}, {King}, {Kirkby},
  {Kulkarni}, {Kumar}, {Lee}, {Lenz}, {Littlefair}, {Ma}, {Macleod},
  {Mastropietro}, {McCully}, {Montagnac}, {Morris}, {Mueller}, {Mumford},
  {Muna}, {Murphy}, {Nelson}, {Nguyen}, {Ninan}, {N{\"o}the}, {Ogaz}, {Oh},
  {Parejko}, {Parley}, {Pascual}, {Patil}, {Patil}, {Plunkett}, {Prochaska},
  {Rastogi}, {Reddy Janga}, {Sabater}, {Sakurikar}, {Seifert}, {Sherbert},
  {Sherwood-Taylor}, {Shih}, {Sick}, {Silbiger}, {Singanamalla}, {Singer},
  {Sladen}, {Sooley}, {Sornarajah}, {Streicher}, {Teuben}, {Thomas},
  {Tremblay}, {Turner}, {Terr{\'o}n}, {van Kerkwijk}, {de la Vega}, {Watkins},
  {Weaver}, {Whitmore}, {Woillez}, {Zabalza}, \& {Contributors}}]{astropy:2018}
{Price-Whelan}, A.~M., {Sip{\H{o}}cz}, B.~M., {G{\"u}nther}, H.~M., {et~al.}
  2018, \aj, 156, 123, \dodoi{10.3847/1538-3881/aabc4f}

\bibitem[{{Purcell} {et~al.}(2007){Purcell}, {Bullock}, \&
  {Zentner}}]{purcell2007}
{Purcell}, C.~W., {Bullock}, J.~S., \& {Zentner}, A.~R. 2007, \apj, 666, 20,
  \dodoi{10.1086/519787}

\bibitem[{{Renzini} \& {Peng}(2015)}]{renzini2015}
{Renzini}, A., \& {Peng}, Y.-j. 2015, \apjl, 801, L29,
  \dodoi{10.1088/2041-8205/801/2/L29}

\bibitem[{{Rodighiero} {et~al.}(2011){Rodighiero}, {Daddi}, {Baronchelli},
  {Cimatti}, {Renzini}, {Aussel}, {Popesso}, {Lutz}, {Andreani}, {Berta},
  {Cava}, {Elbaz}, {Feltre}, {Fontana}, {F{\"o}rster Schreiber},
  {Franceschini}, {Genzel}, {Grazian}, {Gruppioni}, {Ilbert}, {Le Floch},
  {Magdis}, {Magliocchetti}, {Magnelli}, {Maiolino}, {McCracken}, {Nordon},
  {Poglitsch}, {Santini}, {Pozzi}, {Riguccini}, {Tacconi}, {Wuyts}, \&
  {Zamorani}}]{rodighiero2011}
{Rodighiero}, G., {Daddi}, E., {Baronchelli}, I., {et~al.} 2011, \apjl, 739,
  L40, \dodoi{10.1088/2041-8205/739/2/L40}

\bibitem[{{Salim} {et~al.}(2018){Salim}, {Boquien}, \& {Lee}}]{salim2018}
{Salim}, S., {Boquien}, M., \& {Lee}, J.~C. 2018, \apj, 859, 11,
  \dodoi{10.3847/1538-4357/aabf3c}

\bibitem[{{Salim} {et~al.}(2007){Salim}, {Rich}, {Charlot}, {Brinchmann},
  {Johnson}, {Schiminovich}, {Seibert}, {Mallery}, {Heckman}, {Forster},
  {Friedman}, {Martin}, {Morrissey}, {Neff}, {Small}, {Wyder}, {Bianchi},
  {Donas}, {Lee}, {Madore}, {Milliard}, {Szalay}, {Welsh}, \& {Yi}}]{salim2007}
{Salim}, S., {Rich}, R.~M., {Charlot}, S., {et~al.} 2007, \apjs, 173, 267,
  \dodoi{10.1086/519218}

\bibitem[{{Scott}(1992)}]{scott1992}
{Scott}, D.~W. 1992, {Multivariate Density Estimation}

\bibitem[{{Smith} {et~al.}(2024){Smith}, {Fielding}, {Bryan}, {Bennett}, {Kim},
  {Ostriker}, \& {Somerville}}]{smith2024}
{Smith}, M.~C., {Fielding}, D.~B., {Bryan}, G.~L., {et~al.} 2024, arXiv
  e-prints, arXiv:2408.15321, \dodoi{10.48550/arXiv.2408.15321}

\bibitem[{{Speagle} {et~al.}(2014){Speagle}, {Steinhardt}, {Capak}, \&
  {Silverman}}]{speagle2014}
{Speagle}, J.~S., {Steinhardt}, C.~L., {Capak}, P.~L., \& {Silverman}, J.~D.
  2014, \apjs, 214, 15, \dodoi{10.1088/0067-0049/214/2/15}

\bibitem[{{Stierwalt} {et~al.}(2015){Stierwalt}, {Besla}, {Patton}, {Johnson},
  {Kallivayalil}, {Putman}, {Privon}, \& {Ross}}]{stierwalt2015}
{Stierwalt}, S., {Besla}, G., {Patton}, D., {et~al.} 2015, \apj, 805, 2,
  \dodoi{10.1088/0004-637X/805/1/2}

\bibitem[{{Sun} {et~al.}(2023){Sun}, {Leroy}, {Ostriker}, {Meidt},
  {Rosolowsky}, {Schinnerer}, {Wilson}, {Utomo}, {Belfiore}, {Blanc},
  {Emsellem}, {Faesi}, {Groves}, {Hughes}, {Koch}, {Kreckel}, {Liu}, {Pan},
  {Pety}, {Querejeta}, {Razza}, {Saito}, {Sardone}, {Usero}, {Williams},
  {Bigiel}, {Bolatto}, {Chevance}, {Dale}, {Gensior}, {Glover}, {Grasha},
  {Henshaw}, {Jim{\'e}nez-Donaire}, {Klessen}, {Kruijssen}, {Murphy},
  {Neumann}, {Teng}, \& {Thilker}}]{sun2023}
{Sun}, J., {Leroy}, A.~K., {Ostriker}, E.~C., {et~al.} 2023, \apjl, 945, L19,
  \dodoi{10.3847/2041-8213/acbd9c}

\bibitem[{{The pandas development Team}(2024)}]{pandas2022}
{The pandas development Team}. 2024, {pandas-dev/pandas: Pandas}, v2.2.1,
  Zenodo, \dodoi{10.5281/zenodo.3509134}

\bibitem[{{Thorne} {et~al.}(2021){Thorne}, {Robotham}, {Davies}, {Bellstedt},
  {Driver}, {Bravo}, {Bremer}, {Holwerda}, {Hopkins}, {Lagos}, {Phillipps},
  {Siudek}, {Taylor}, \& {Wright}}]{thorne2021}
{Thorne}, J.~E., {Robotham}, A. S.~G., {Davies}, L. J.~M., {et~al.} 2021,
  \mnras, 505, 540, \dodoi{10.1093/mnras/stab1294}

\bibitem[{{Tomczak} {et~al.}(2016){Tomczak}, {Quadri}, {Tran}, {Labb{\'e}},
  {Straatman}, {Papovich}, {Glazebrook}, {Allen}, {Brammer}, {Cowley},
  {Dickinson}, {Elbaz}, {Inami}, {Kacprzak}, {Morrison}, {Nanayakkara},
  {Persson}, {Rees}, {Salmon}, {Schreiber}, {Spitler}, \&
  {Whitaker}}]{tomczak2016}
{Tomczak}, A.~R., {Quadri}, R.~F., {Tran}, K.-V.~H., {et~al.} 2016, \apj, 817,
  118, \dodoi{10.3847/0004-637X/817/2/118}

\bibitem[{{Tully} \& {Fisher}(1977)}]{tully1977}
{Tully}, R.~B., \& {Fisher}, J.~R. 1977, \aap, 54, 661

\bibitem[{Van Der~Walt {et~al.}(2011)Van Der~Walt, Colbert, \&
  Varoquaux}]{van2011numpy}
Van Der~Walt, S., Colbert, S.~C., \& Varoquaux, G. 2011, Computing in Science
  \& Engineering, 13, 22

\bibitem[{{Virtanen} {et~al.}(2020){Virtanen}, {Gommers}, {Oliphant},
  {Haberland}, {Reddy}, {Cournapeau}, {Burovski}, {Peterson}, {Weckesser},
  {Bright}, {van der Walt}, {Brett}, {Wilson}, {Millman}, {Mayorov}, {Nelson},
  {Jones}, {Kern}, {Larson}, {Carey}, {Polat}, {Feng}, {Moore}, {VanderPlas},
  {Laxalde}, {Perktold}, {Cimrman}, {Henriksen}, {Quintero}, {Harris},
  {Archibald}, {Ribeiro}, {Pedregosa}, {van Mulbregt}, \& {SciPy 1. 0
  Contributors}}]{scipy2020}
{Virtanen}, P., {Gommers}, R., {Oliphant}, T.~E., {et~al.} 2020, Nature
  Methods, 17, 261, \dodoi{10.1038/s41592-019-0686-2}

\bibitem[{{Wang} {et~al.}(2015){Wang}, {Dutton}, {Stinson}, {Macci{\`o}},
  {Penzo}, {Kang}, {Keller}, \& {Wadsley}}]{wang2015}
{Wang}, L., {Dutton}, A.~A., {Stinson}, G.~S., {et~al.} 2015, \mnras, 454, 83,
  \dodoi{10.1093/mnras/stv1937}

\bibitem[{{Wang} {et~al.}(2018){Wang}, {Pearce}, {Knebe}, {Yepes}, {Cui},
  {Power}, {Arth}, {Gottl{\"o}ber}, {De Petris}, {Brown}, \& {Feng}}]{wang2018}
{Wang}, Y., {Pearce}, F., {Knebe}, A., {et~al.} 2018, \apj, 868, 130,
  \dodoi{10.3847/1538-4357/aae52e}

\bibitem[{{Wang} {et~al.}(2024){Wang}, {Nadler}, {Mao}, {Wechsler}, {Abel},
  {Behroozi}, {Geha}, {Asali}, {de los Reyes}, {Kado-Fong}, {Kallivayalil},
  {Tollerud}, {Weiner}, \& {Wu}}]{sagav}
{Wang}, Y., {Nadler}, E.~O., {Mao}, Y.-Y., {et~al.} 2024, arXiv e-prints,
  arXiv:2404.14500, \dodoi{10.48550/arXiv.2404.14500}

\bibitem[{{Whitaker} {et~al.}(2012){Whitaker}, {van Dokkum}, {Brammer}, \&
  {Franx}}]{whitaker2012}
{Whitaker}, K.~E., {van Dokkum}, P.~G., {Brammer}, G., \& {Franx}, M. 2012,
  \apjl, 754, L29, \dodoi{10.1088/2041-8205/754/2/L29}

\bibitem[{{Whitaker} {et~al.}(2014){Whitaker}, {Franx}, {Leja}, {van Dokkum},
  {Henry}, {Skelton}, {Fumagalli}, {Momcheva}, {Brammer}, {Labb{\'e}},
  {Nelson}, \& {Rigby}}]{whitaker2014}
{Whitaker}, K.~E., {Franx}, M., {Leja}, J., {et~al.} 2014, \apj, 795, 104,
  \dodoi{10.1088/0004-637X/795/2/104}

\bibitem[{{Wuyts} {et~al.}(2011){Wuyts}, {F{\"o}rster Schreiber}, {van der
  Wel}, {Magnelli}, {Guo}, {Genzel}, {Lutz}, {Aussel}, {Barro}, {Berta},
  {Cava}, {Graci{\'a}-Carpio}, {Hathi}, {Huang}, {Kocevski}, {Koekemoer},
  {Lee}, {Le Floc'h}, {McGrath}, {Nordon}, {Popesso}, {Pozzi}, {Riguccini},
  {Rodighiero}, {Saintonge}, \& {Tacconi}}]{wuyts2011}
{Wuyts}, S., {F{\"o}rster Schreiber}, N.~M., {van der Wel}, A., {et~al.} 2011,
  \apj, 742, 96, \dodoi{10.1088/0004-637X/742/2/96}

\bibitem[{{Yao} {et~al.}(2022){Yao}, {Chen}, {Liu}, {Chen}, {Lin}, {Zhang},
  {Gao}, \& {Kong}}]{yao2022}
{Yao}, Y., {Chen}, G., {Liu}, H., {et~al.} 2022, \aap, 661, A112,
  \dodoi{10.1051/0004-6361/202243104}

\end{thebibliography}

\appendix{
\vspace{-15pt}
\section{The Retrogressed \sagabg{} Reference Sample}\label{s:appendix:fits}
In the main text of this work, we showed only the best-fit power-laws associated with the 
underlying \sagabg{} reference sample retrogressed to each redshift bin (see \autoref{f:SFS_datalfits}). In \autoref{f:fullthiswork_sfs}, we show the distribution of the 
retrogressed \sagabg{} sample without applying the observability criteria used to produce \autoref{f:SFS_evolution}. As shown in \autoref{f:fullthiswork_sfs} 
and implied by the fits presented in \autoref{f:SFS_evolution}, we do not find strong evidence for a 
significant evolution in the slope or intrinsic scatter of the SFS over this 
redshift range. For accessibility, we include a table of the fits 
tabulated in \autoref{f:fullthiswork_sfs} in \autoref{t:sfs_fits}.

\begin{figure}[b]
    \centering
    \includegraphics[width=.87\linewidth]{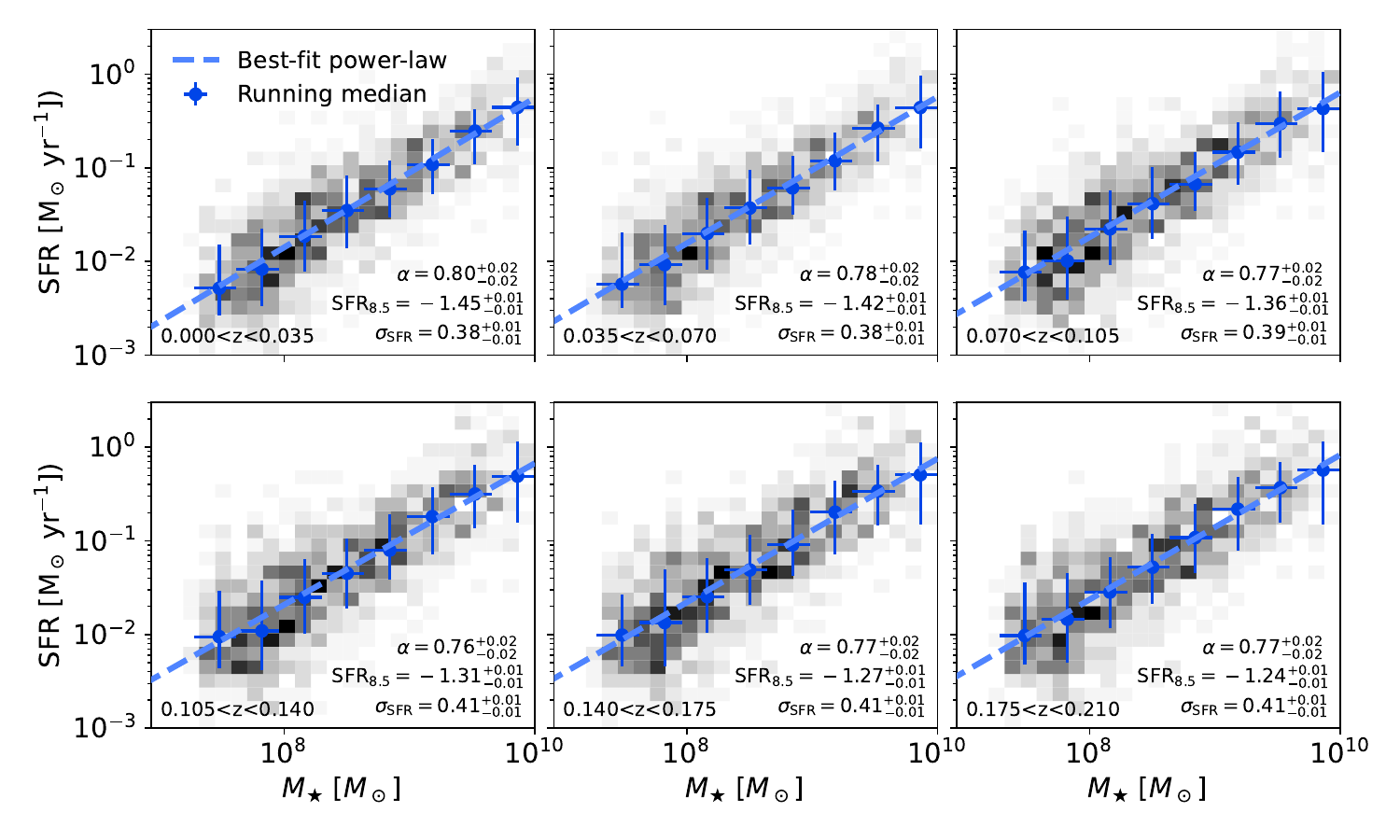}
    \caption{
        The \sagabg{} SFS in the same redshift bins considered in \autoref{f:SFS_datalfits} without imposing our observability criteria.
        The retrogressed sample is shown in grey, while the
        best-fit power-law and running median are shown in blue (where the 
        errorbars indicate the 16\tth{} and 84\tth{} percentiles). The best-fit parameters of \autoref{e:plaw} are again given for each redshift bin.
    }
    \label{f:fullthiswork_sfs}
\end{figure}

\begin{deluxetable}{cccccccccc}[b]
\setlength{\tabcolsep}{3pt}
\tablecaption{Power-law fits to the underlying \sagabg{} SFS}\label{t:sfs_fits}
\tablewidth{0pt}
\tablehead{
\colhead{Redshift} & \colhead{$q_{16}(\alpha)$} & \colhead{$\mathbf{q_{50}(\alpha)}$} & \colhead{$q_{84}(\alpha)$} & \colhead{$q_{16}({\rm SFR}_{8.5})$} & \colhead{$\mathbf{q_{50}({\rm SFR}_{8.5})}$} & \colhead{$q_{84}({\rm SFR}_{8.5})$} & \colhead{$q_{16}(\sigma_{\rm SFR})$} & \colhead{$\mathbf{q_{50}(\sigma_{\rm SFR})}$} & \colhead{$q_{84}(\sigma_{\rm SFR})$}\\
\colhead{} & \colhead{} & \colhead{} & \colhead{} & \colhead{$\rm [M_\odot\ yr^{-1}]$} & \colhead{$\rm [M_\odot\ yr^{-1}]$} & \colhead{$\rm [M_\odot\ yr^{-1}]$} & \colhead{$\rm [M_\odot\ yr^{-1}]$} & \colhead{$\rm [M_\odot\ yr^{-1}]$} & \colhead{$\rm [M_\odot\ yr^{-1}]$}
}
\startdata
$0.000<z<0.035$ & 0.778 & \textbf{0.796} & 0.813 & -1.463 & \textbf{-1.452} & -1.440 & 0.367 & \textbf{0.376} & 0.384 \\
$0.035<z<0.070$ & 0.762 & \textbf{0.781} & 0.800 & -1.426 & \textbf{-1.415} & -1.403 & 0.376 & \textbf{0.385} & 0.394 \\
$0.070<z<0.105$ & 0.755 & \textbf{0.774} & 0.792 & -1.372 & \textbf{-1.359} & -1.347 & 0.382 & \textbf{0.391} & 0.399 \\
$0.105<z<0.140$ & 0.736 & \textbf{0.756} & 0.776 & -1.319 & \textbf{-1.307} & -1.293 & 0.399 & \textbf{0.408} & 0.418 \\
$0.140<z<0.175$ & 0.751 & \textbf{0.769} & 0.790 & -1.289 & \textbf{-1.274} & -1.260 & 0.397 & \textbf{0.407} & 0.418 \\
$0.175<z<0.210$ & 0.748 & \textbf{0.769} & 0.789 & -1.249 & \textbf{-1.234} & -1.219 & 0.406 & \textbf{0.416} & 0.427 \\
\enddata
\tablenotetext{}{In each column, $q_{XX}(Y)$ refers to the $XX$\textsuperscript{th} percentile of the posterior sample over the
inferred parameter $Y$. The median parameter estimates are highlighted in bold for clarity.}
\end{deluxetable}
}

\section{\rrr{Validating Model Fits}}\label{s:appendix:modelcomparison}

\begin{figure}[t]
  \centering
  \includegraphics[width=\linewidth]{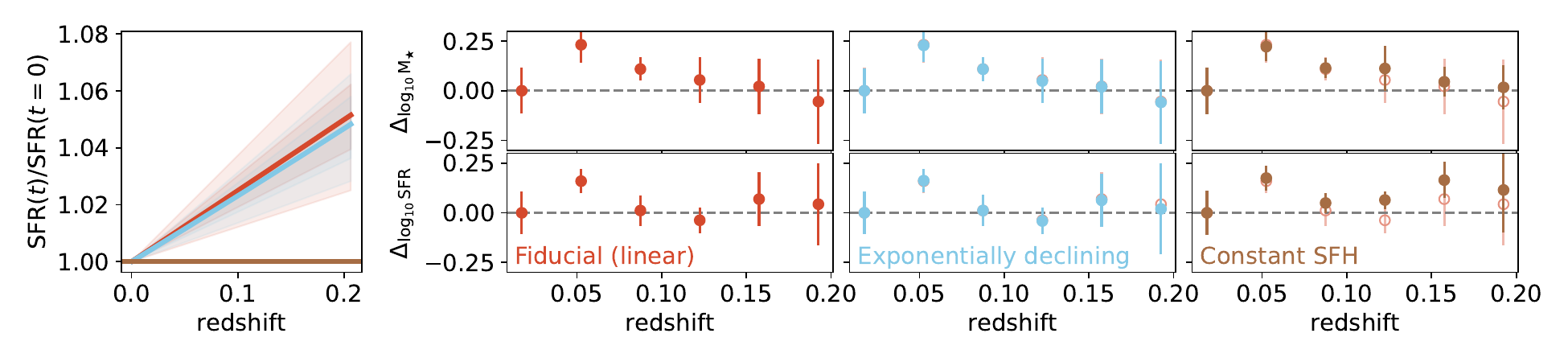}
  \caption{
      \rrr{\textit{Left:} redshift versus the physical change in average SFR for our fiducial model (red), an exponentially declining $\tau$ model (blue), and a constant SFH (brown). Our best-fit $\tau$ produces a relationship between SFR and lookback time that is statistically indistinguishable from our fiducial linear model. \textit{Right:} for each model, we show the difference in  the median stellar mass (top) and star formation rate (bottom) between the observed \sagabg{} sample and observable retrogressed model prediction. We define $\Delta {\log_{10}\mathcal{X}}$ to be $\Delta {\log_{10}\mathcal{X}} \equiv \log_{10}\mathcal{X}_{\rm obs} - \log_{10}\mathcal{X}_{\rm model}$ for some observed proeprty $\mathcal{X}$ of the galaxy sample. For our non-fiducial models, we underlay the fiducial model performance as unfilled red errorbars.}
  }
  \label{f:appendix_comparison}
\end{figure}

\rrr{We adopt a simple model for the recent average change in the SFS as a function of lookback time. To check if this assumption affects our inferred shift in the SFS, we compare to 
an exponentially declining $\tau$ model (wherein $ {\rm SFR}(t) \propto \exp({t/\tau})$ for some characteristic timescale $\tau$) and a static SFR model. We show the results of this exercise in \autoref{f:appendix_comparison}; the model performance is quantified by a comparison between the median stellar mass and SFR of the observed \sagabg{} sample and the observable retrogressed model predictions. The 68\% confidence intervals are shown as errorbars in all panels. We find that our fiducial model and 
the exponentially declining $\tau$ model produce statistically indistinguishable results, and that assuming a constant star formation history underpredicts the relation between SFR and redshift in the \sagabg{} sample.} 

\end{document}